# An End-to-End Differentiable Forward Model for High-Energy Diffraction Microscopy

Hemant Sharma ✉ [a], Nina Andrejevic[a], Simon Zhang[a], and Mathew Cherukara[a]

[a]Advanced Photon Source, Argonne National Laboratory, 9700 South Cass Avenue, Lemont, IL 60439, USA.

**Synopsis**

The first end-to-end differentiable forward model for high-energy diffraction microscopy—spanning far-field, near-field, and point-focused geometries with pixel-exact agreement against the established reference simulators—enabling gradient-based joint refinement of detector geometry and per-grain orientation, strain, and position.

**Abstract**

High-Energy Diffraction Microscopy (HEDM) recovers crystallographic orientation, strain, and grain position from rotating-crystal X-ray diffraction patterns. Existing forward models in far-field (FF), near-field (NF), and point-focused (pf) HEDM are not differentiable, which forecloses gradient-based joint parameter refinement, physics-informed regularisation, and Bayesian uncertainty quantification. We present the first end-to-end differentiable HEDM forward model covering all three geometries, implemented in PyTorch with pixel-exact agreement against the established MIDAS reference simulators (162/162 FF, 2304/2304 NF including a non-zero detector-tilt sweep, and 1088/1096 pf-HEDM spots matched). Three demonstrations validate the framework: joint orientation-strain-position recovery in NF-HEDM at $\sim 6\,\mathrm{nm}$ precision; round-trip refinement on a real 214-grain $\alpha$-Ti FF-HEDM dataset reaching 100% grain recovery from a 1.5° initial perturbation with residuals matching the production fit to 0.3%; and joint refinement of all per-detector geometry parameters and per-grain state on a synthetic four-panel FF-HEDM setup, recovering panel rotations about the beam axis to $\sim 10\,\mu$rad and a global rotation-axis wedge to $\sim 26\,\mu$rad. The framework is released as the open-source `midas-diffract` package (`pip install midas-diffract`).

## 1 Introduction

High-Energy Diffraction Microscopy (HEDM) is the workhorse technique for three-dimensional non-destructive characterisation of polycrystalline metals at synchrotron sources worldwide (Poulsen, 2004; Suter *et al.*, 2006; Sharma *et al.*, 2026*a*; Sharma *et al.*, 2026*b*). The three variants; far-field (FF), near-field (NF), and point-focused (pf-HEDM) or scanning three-dimensional X-ray diffraction (s3DXRD) each measure diffraction patterns from a rotating polycrystalline sample and solve an inverse problem to recover grain-level quantities: crystallographic orientation, lattice strain, grain position, morphology, and intragranular deformation.

Forward modelling is the common substrate of nearly every quantitative materials characterisation technique. Rietveld refinement of powder diffraction (Rietveld, 1969), dynamical image simulation in high-resolution and four-dimensional scanning transmission electron microscopy, electron-backscatter- diffraction pattern matching, single-crystal structure refinement, atom-probe field-ion reconstruction, and ptychographic phase retrieval all reduce to the same template: a physics-based map from a parameterised microstructural state to a measured signal, inverted by minimising a residual against data. What varies across these communities is not the existence of the forward model but the cost of its derivatives—each technique has historically accumulated bespoke analytic Jacobians, finite-difference approximations, or task-specific gradient codes. Automatic differentiation removes that bottleneck uniformly, and HEDM is one of the last major synchrotron diffraction techniques where this transition has not yet been made.

At the heart of every HEDM reconstruction pipeline is a *forward model*: a function that maps grain state (orientation, strain, position) and instrument state (detector geometry, beam parameters, sample environment) to simulated detector signal. All widely-used HEDM codes—MIDAS, HEXRD, ImageD11, hexomap—implement this forward model in C or C++, optimised for throughput. The C implementation is highly performant and battle-tested against two decades of experimental data, but it carries a structural limitation: *the forward model is not differentiable with respect to its inputs.* Gradients of the simulated signal with respect to orientation, strain, or geometry cannot be computed without finite-difference approximation, which is expensive and noisy.

This non-differentiability forecloses an entire class of modern inference methods:

- Gradient-based *joint* optimisation of crystallographic and instrument parameters, which can escape the local minima that sequential optimisation gets stuck in.
- Regularised reconstruction with both first-principles physics constraints (crystal-symmetry / fundamental-zone projection of orientations, positivity and normalisation of density and orientation distribution functions) and microstructure priors (total variation, Markov random fields, manifold priors that encode the piecewise quasi-homogeneous nature of polycrystalline grain interiors), where the regulariser gradient must flow back through the forward model.
- Bayesian uncertainty quantification via Hamiltonian Monte Carlo or variational inference, both of which require exact gradients of the likelihood with respect to model parameters.
- End-to-end coupling to differentiable simulation frameworks (e.g., JAX-FEM for crystal plasticity), enabling direct comparison of experimental HEDM output with simulated material response under load.

- End-to-end learned reconstruction pipelines where the physics-based forward model serves as a constraint layer inside a larger neural-network architecture.

Learned surrogates for the forward model (Liu *et al.*, 2022; Neural Architecture Codesign authors, 2023) can restore differentiability but introduce approximation error that is difficult to characterise and incompatible with the physical conventions used by the rest of the HEDM toolchain. The open problem so far is whether HEDM can be made differentiable *while preserving pixel-exact agreement with the validated C physics.* Concurrent 2025 work in electron diffraction (Fang *et al.*, 2025) has explored differentiable physics models in that modality; to our knowledge, no existing work achieves the same for high-energy X-ray diffraction of polycrystals, which is the subject of this paper.

This paper closes that gap. We present an end-to-end differentiable HEDM forward model implemented in PyTorch, supporting FF, NF, and pf-HEDM geometries, with the following contributions:

1. **End-to-end differentiability across all three HEDM modalities** (FF, NF, pf), in a single unified PyTorch implementation. No prior open-source HEDM forward simulator (including the most recent, `xrd_simulator` (Henningsson & Hall, 2023), which uses PyTorch as a tensor backend without enabling autograd) exposes gradients through the diffraction physics. Differentiable surrogates (Liu *et al.*, 2022; Neural Architecture Codesign authors, 2023) restore gradients only through learned approximations to the physics.

2. **Pixel-exact agreement with the established MIDAS reference forward simulators**, validated to floating-point precision: 162/162 FF spots matched with sub-$10^{-3}$ pixel residuals; 2304/2304 NF pixels matched identically, preserved under a sweep of non-zero detector tilts up to 1° per axis; 1088/1096 pf-HEDM spots matched, with 8 unmatched reflections at the near-axis branch-selection boundary. Section 5 documents the agreement and the origin of every residual; the MIDAS reference simulators are described in supplementary SS1.

3. **Real-data round-trip on a 214-grain $\alpha$-Ti FF-HEDM reconstruction at the 1-ID-E beamline**: forward-model residuals match the production refined fit to within 0.3%, perturb-and-recover on the actual measured spots achieves 100% grain recovery from a 1.5° initial perturbation, and the real-data basin of convergence matches the synthetic measurement.

4. **Joint refinement of multi-panel detector geometry and per-grain state** in a single gradient flow on a four-panel synthetic FF setup, including the global rotation-axis wedge – a regime that existing FF pipelines handle one panel at a time.

5. **Joint orientation-strain-position recovery in NF-HEDM** to $\sim 6\,\mathrm{nm}$ precision in a single gradient flow, a continuous-coordinate capability that voxel-grid HEDM pipelines cannot straightforwardly deliver.

6. **Open-source software release** as `midas-diffract` v0.1.0 on PyPI, with Zenodo DOI and reproducibility scripts for every figure in this paper.

The resulting framework is the methodological foundation for a series of follow-on capabilities; sub-voxel grain mixtures, physics-informed regularization, temporal 4D-HEDM tracking, crystal-plasticity coupling, and Bayesian HEDM each of which requires end-to-end differentiability and each of which is the subject of a forthcoming paper.

# 2 Related Work

## 2.1 HEDM Forward Models in the Literature

The physical basis of HEDM forward simulation was established by the foundational work of Poulsen (Poulsen, 2004) for 3DXRD and by Suter *et al.* (Suter *et al.*, 2006) for the near-field variant. Production implementations followed in several community codes: MIDAS (Sharma *et al.*, 2026*a*; Sharma *et al.*, 2026*b*) at APS, HEXRD (Boyce & Bernier, 2013; Bernier *et al.*, 2011; Pagan *et al.*, 2020) at LLNL/CHESS, ImageD11 (Wright *et al.*, 2024) at ESRF for grain indexing, and the FABLE s3DXRD scanning-diffraction toolkit (Henningsson & Hendriks, 2024). All of these implementations share a common architectural choice: the forward model is written in C, C++, or Fortran for throughput, and gradients with respect to grain or instrument parameters are either computed by finite difference (expensive, noisy) or not at all. Recent work on forward-model-based grain reconstruction for diffraction contrast tomography (Henningsson & Hall, 2025) continues in the same non-differentiable tradition.

The most recent open-source HEDM forward simulator is `xrd_simulator` (Henningsson & Hall, 2023), which builds on the tetrahedral-mesh lineage of Wong *et al.* (Wong *et al.*, 2013) and supports arbitrary rigid-body sample motions, multi-phase polycrystals with per-element orientation and strain, and convex-polyhedral beams intersected with mesh elements to model partial illumination intensities and spatially extended peak shapes. With respect to the Laue / UB physics that governs peak positions, our forward model is equivalent: seeding one "grain" per mesh element reproduces identical diffraction vectors and detector positions, and our model likewise admits arbitrary sample rotations and translations and per-grain intragranular variation. The essential distinctions are orthogonal: (i) our forward model is end-to-end differentiable, which `xrd_simulator` is not; (ii) we cover NF-HEDM as well as FF and pf-HEDM, which `xrd_simulator` does not; and (iii) our intensity model follows the point-source convention common to production HEDM pipelines (MIDAS, HEXRD), trading the volume-intersection peak-shape fidelity of `xrd_simulator` for pixel-exact agreement with the canonical C reference against which community workflows have been validated. A differentiable volume-intersection intensity layer is a natural extension and is outside the scope of this paper.

## 2.2 Learned Surrogates for HEDM

The BraggNN family (Liu *et al.*, 2022; Neural Architecture Codesign authors, 2023) replaces the Bragg-peak localisation step of the HEDM pipeline with a convolutional neural network, achieving two orders of magnitude speed-up over iterative pseudo-Voigt fitting and improving downstream grain position accuracy against NF-HEDM ground truth. Learned surrogates of this type are fully differentiable by construction (they are neural networks), but they sacrifice the pixel-exact physics against which the community's reconstruction pipelines have been validated: the surrogate's forward mapping is an approximation of the true diffraction physics, with an approximation error that is empirical rather than mathematically bounded. Our work is orthogonal to, and composable with, this line: the differentiable forward model we present can be used as a physics constraint layer inside a BraggNN-style pipeline, retaining end-to-end differentiability while restoring the exact forward physics wherever it is preferred.

## 2.3 Differentiable Physics in Adjacent Imaging Fields

The methodological closest analogue to our work is Kandel *et al.*'s use of automatic differentiation as a general framework for ptychographic reconstruction (Kandel *et al.*, 2019), subsequently extended to X-ray nanotomography (Du *et al.*, 2020). Both establish that a physics-based forward model, expressed in an autograd-capable language, can be used as the backbone of gradient-based inverse problems in X-ray imaging; the present paper does the same for HEDM. In cryo-EM, CryoDRGN (Zhong *et al.*, 2021) pioneered end-to-end differentiable reconstruction of heterogeneous structures with PyTorch. Neural-network-augmented ptychographic reconstruction has progressed from off-line training (Cherukara *et al.*, 2020) to real-time edge deployment (Babu *et al.*, 2023). The broader machinery of differentiable rendering in computer vision, soft rasterizers (Liu *et al.*, 2019) and the PyTorch3D framework (Ravi *et al.*, 2020)—provides the infrastructural template for treating a physics simulator as a differentiable function inside a larger optimisation loop.

Within the synchrotron X-ray community itself, automatic differentiation has matured on three fronts that sit closer to the present work than computer-vision rendering. For X-ray fluorescence analysis, Jiang *et al.* (Jiang *et al.*, 2026) have released `MapsTorch`, an AD-based spectrum-fitting framework that replaces the long-standing iterative XRF analysis pipeline with a differentiable computation graph and consistently matches or exceeds the fit quality of the established `NLopt` tooling. For X-ray ptychography, Wu *et al.* (Wu *et al.*, 2024) demonstrated that AD-based reconstruction substantially improves dose efficiency, and the recent open-source `Pty-Chi` framework (Du *et al.*, 2025) builds modern ptychography directly on PyTorch's autograd with multi-GPU support and a modular probe / object / multislice abstraction. These efforts establish AD as a credible production tool for synchrotron data analysis, but each targets a *coherent-imaging* inverse problem: recovery of a continuous complex-valued object or spectrum from a smooth diffraction or fluorescence intensity field. HEDM presents a structurally distinct challenge—discrete Bragg reflections distributed over an $\omega$-sweep, parameterised by per-grain orientation, position, and strain tensors.

The forward model is geometric and combinatorial (spot indexing, branch selection, fundamental-zone symmetry, soft rasterisation onto a pixel grid) before it is photometric. Differentiable ptychography tooling does not transfer directly: HEDM requires a forward model whose derivatives carry through the indexing geometry, which is the contribution of this paper.

### 2.4 Adjacent-Modality Differentiable Diffraction

Concurrent 2025 work by Fang *et al.* (Fang *et al.*, 2025) develops a hybrid physics/machine-learning forward model for quantitative 3D *electron* diffraction refinements. Both that work and the present paper share the philosophical position that a physics forward model should be differentiable end-to-end; the modalities, geometries, and sample regimes addressed are nevertheless distinct. Electron diffraction operates at very different wavelengths, cross-sections, and sample thicknesses, and the relevant reconstruction problems (crystal-structure refinement, dynamical scattering) do not overlap with the polycrystalline grain-mapping regime of HEDM. To our knowledge, no published work addresses end-to-end differentiable forward modelling for high-energy X-ray diffraction of polycrystals.

### 2.5 Gap This Paper Closes

Across the literature surveyed above, no existing HEDM implementation provides end-to-end differentiability with pixel-exact agreement against a validated reference forward simulator. BraggNN-style surrogates are differentiable but approximate; existing C simulators and their most recent Python successor `xrd_simulator` (Henningsson & Hall, 2023) are physically rich but not differentiable; adjacent-modality differentiable physics does not cover the HEDM regime. The present work closes that gap for all three principal HEDM geometries (FF, NF, pf-HEDM) in a single unified framework (Figure 1).

## 3 Reference Forward Simulator

The differentiable forward model introduced in Section 4 is validated against the two long-standing reference forward simulators distributed with MIDAS: `ForwardSimulationCompressed` for FF and pf-HEDM, and `simulateNF` for NF-HEDM. Both have been in production use at APS and partner facilities for over a decade, generating synthetic benchmarks for indexing and fitting code and diagnosing beamline issues by comparing simulated against measured patterns. Their battle-tested status makes them the natural reference: any claim of "pixel-exact agreement" in this paper is a claim against these two codes under identical inputs. A self-contained description of both simulators – inputs, physics conventions, parallelism, output layout – is given in supplementary SS1.

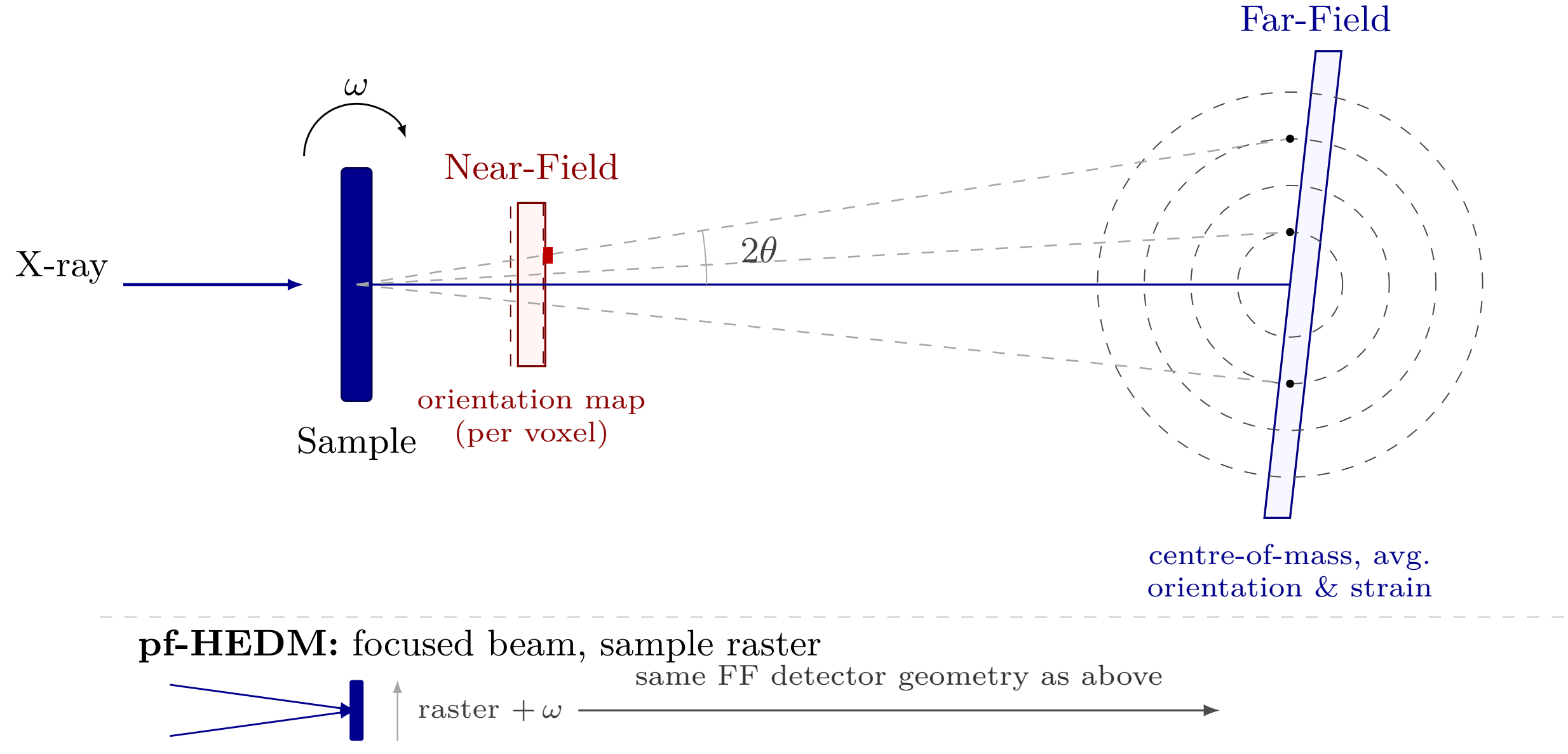

Figure 1: HEDM geometries unified by this work, drawn on a single beam axis (adapted from the layout convention used for HEDM at the Advanced Photon Source). **(top)** The X-ray beam enters from the left, passes through the rotating sample (rotation about $\omega$), and exits to the right. The **Near-Field (NF)** detector is positioned close to the sample ($L_{sd} \approx 5\,\mathrm{mm}$) so the projection of grain shape onto the detector is resolvable; the standard NF reconstruction product is a voxelised grain orientation map. The **Far-Field (FF)** detector is positioned $\sim 1\,\mathrm{m}$ downstream; the Debye–Scherrer rings host sparse Bragg spots, and per-grain indexing recovers centre-of-mass position, average orientation, and average strain. **(bottom) Point-focused (pf-HEDM)** uses the same FF detector geometry, but with the beam focused down to a small spot and the sample rastered through it while rotating, giving sub-grain spatial resolution from per-scan diffraction patterns. The differentiable forward model presented in this paper covers all three geometries within a single unified PyTorch module.

# 4 Differentiable Forward Model

## 4.1 Architectural Overview

We implement the differentiable HEDM forward model as a single PyTorch (Paszke *et al.*, 2019) `nn.Module` class, `HEDMForwardModel`, that handles all three geometries (FF, NF, pf-HEDM) through *configuration* rather than subclassing. The physical core; orientation to reciprocal-space vectors, omega solver, azimuthal angle computation, and detector projection is identical across modalities. Modality differences (single vs. multi-distance detector layout, beam translation for pf-HEDM, angular-coordinate vs. image output) enter through three data classes: `HEDMGeometry` (detector distances, beam-centre pixel coordinates, pixel size, omega sweep, wavelength, detector dimensions); `ScanConfig` (per-scan beam positions and beam size for pf-HEDM); and `TriVoxelConfig` (triangular voxel grid parameters for NF-HEDM). This unified architecture makes cross-modality validation straightforward: the same omega quadratic solver must produce pixel-exact agreement against both `ForwardSimulationCompressed` (FF and pf) and `simulateNF` (NF), and any divergence can be traced to a single codepath.

Figure 2 summarises the pipeline. The forward pass takes as inputs a batch of grain (or voxel) descriptors, Euler angles, grain positions, and optional lattice parameters plus a precomputed list of integer Miller indices and their nominal Bragg angles, and returns a `SpotDescriptors` structure holding the predicted spot coordinates, validity mask, and (for NF) a synthesised detector image volume. Every arithmetic operation in the forward pass is a differentiable PyTorch primitive, so gradients of any scalar loss on the output flow back through the full physics.

## 4.2 Pipeline Stages

### 4.2.1 Orientation to rotation matrix.

Grain or voxel orientations are specified as ZXZ Euler angles $(\varphi_1, \Phi, \varphi_2)$ and converted to rotation matrices via the standard ZXZ composition. Rather than taking the Euler representation as ground truth, the resulting matrix is projected back onto the special orthogonal group $SO(3)$ using one Newton–Schulz iteration of the polar decomposition,

$$\mathbf{Q}_{k+1} = \tfrac{1}{2}\,\mathbf{Q}_k(3\mathbf{I} - \mathbf{Q}_k^\top \mathbf{Q}_k), \tag{1}$$

which converges quadratically for near-orthogonal input and reduces residual deviation from $SO(3)$ below $10^{-14}$ after a single step for typical Euler-to-matrix output. Newton–Schulz is preferred over SVD-based projection because SVD is not differentiable at repeated singular values, whereas (1) is a sequence of matrix products with well-behaved gradients everywhere. A determinant-sign guard ensures the result is a proper rotation.

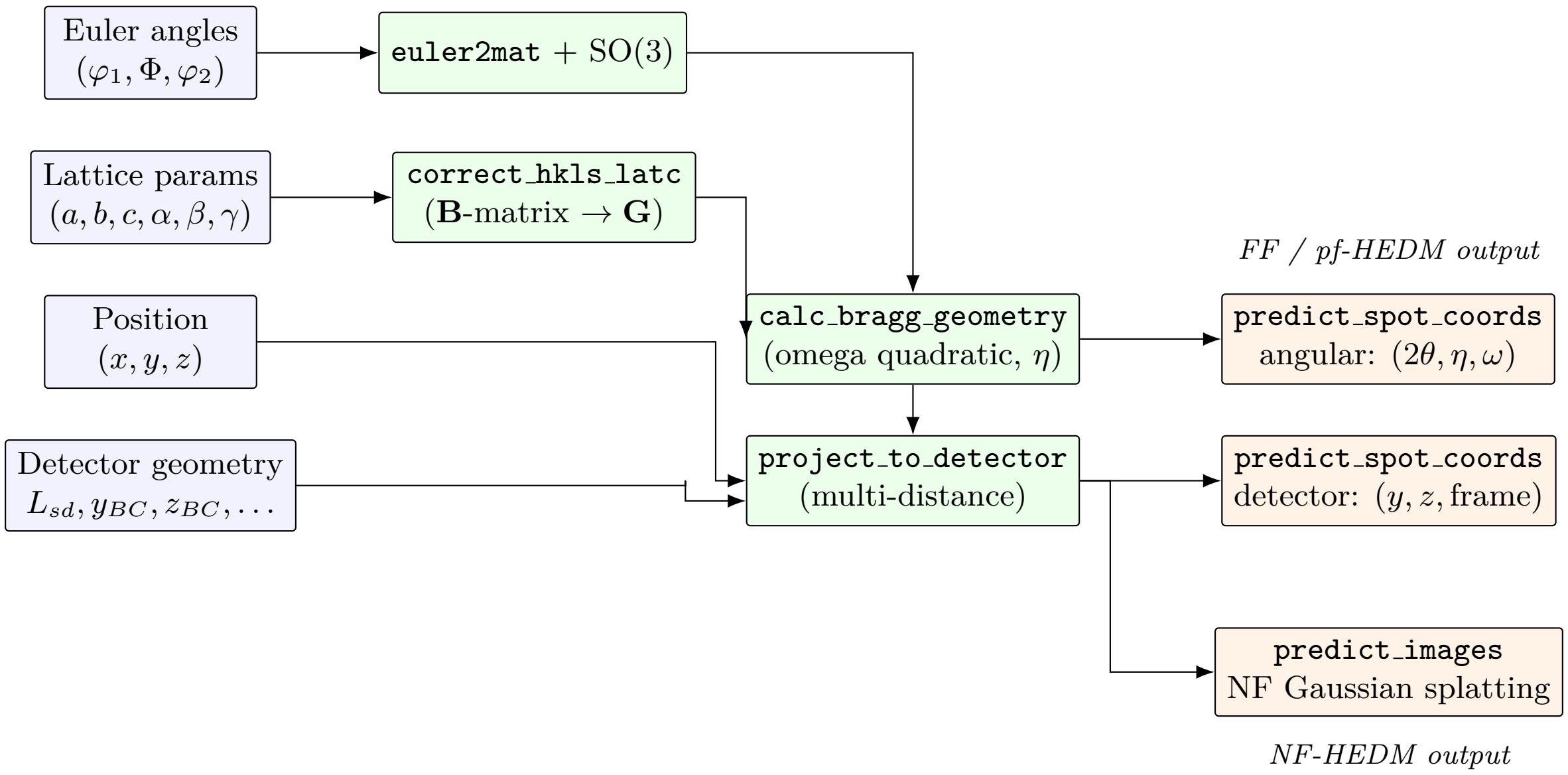

Figure 2: Pipeline of the differentiable HEDM forward model. Grain state (orientation, position, optional lattice parameters) and detector geometry enter a shared physics core (`euler2mat` + SO(3) projection, $\mathbf{B}$-matrix strain, omega quadratic solver, position-dependent multi-distance projection) before branching into mode-specific output heads: angular coordinates for FF and pf-HEDM, Gaussian-splatted detector volume for NF-HEDM. Every operation is a differentiable PyTorch primitive, so scalar losses on any output produce analytic gradients with respect to every input.

#### 4.2.2 Strain via lattice-parameter mapping or explicit strain tensor.

Per-grain strain can be specified through either of two equivalent channels, both of which participate in the autograd graph and may be used together or separately. The first is a deviation in the six lattice parameters $(a, b, c, \alpha, \beta, \gamma)$: the model constructs the reciprocal-lattice $\mathbf{B}_0$ matrix from the standard crystallographic expressions for reciprocal-lattice angles $\gamma'$, $\beta'$ and reciprocal lengths $a'$, $b'$, $c'$, maps integer Miller indices to Cartesian reciprocal-space vectors via $\mathbf{G}_{\text{cart}} = \mathbf{B}_0\,\mathbf{h}$, and computes the corresponding Bragg angles via Bragg's law. The second channel accepts a six-component symmetric infinitesimal strain tensor $\boldsymbol{\epsilon} = [\epsilon_{11}, \epsilon_{12}, \epsilon_{13}, \epsilon_{22}, \epsilon_{23}, \epsilon_{33}]$ expressed in the crystal frame, and applies the Finger-style deformation $\mathbf{B} = (\mathbf{I} + \boldsymbol{\epsilon})^{-1}\,\mathbf{B}_0$ before the HKL contraction. A sample-frame strain tensor is converted to crystal coordinates by a single helper, `rotate_strain_sample_to_crystal`, implementing $\boldsymbol{\epsilon}_c = \mathbf{U}^\top \boldsymbol{\epsilon}_s \mathbf{U}$. Both channels are direct ports of `CorrectHKLsLatC` and `CorrectHKLsLatCEpsilon` from `FF_HEDM/src/ForwardSimulationCompressed.c`; faithful correspondence was verified against the C code line by line and numerically at the precision level used for the validation runs reported in Section 5. Exposing both paths matches the MIDAS production convention, where lattice-parameter refinement is used for orientation-free strain states and an explicit strain tensor is preferred whenever the reconstruction loop must couple to an external mechanical simulation (a crystal-plasticity solver, for instance) that supplies strain directly.

### 4.2.3 Bragg geometry and the omega quadratic.

For each orientation-rotated **G**-vector, we solve the rotating-crystal diffraction condition

$$-G_x \cos\omega + G_y \sin\omega = |\mathbf{G}| \sin\theta \tag{2}$$

for $\omega$, yielding up to two physical solutions per reflection per omega sweep. The solution is obtained by reducing the diffraction equation to a quadratic in $\cos\omega$, solving analytically, and selecting the branch that satisfies the original equation to within a small tolerance. When $|G_y|$ falls below a fixed threshold ($10^{-12}$, matching the C convention), the quadratic degenerates and the solution reduces to $\omega = \pm\arccos(-|\mathbf{G}|\sin\theta/G_x)$; the model handles this case by `torch.where` selecting between the two codepaths rather than branching in Python, preserving differentiability across the boundary. The implementation ports the quadratic solver from `NF_HEDM/src/CalcDiffractionSpots.c:87--183`. Once $\omega$ is in hand, the azimuthal angle $\eta$ is computed from the omega-rotated **G**-vector components in the laboratory frame, and validity masks are assembled from the quadratic discriminant, the $\cos\omega \in [-1, 1]$ constraint, and the user-provided minimum $\eta$ angle.

### 4.2.4 Position-dependent detector projection.

Grains (FF/pf) and voxels (NF) are not at the origin of the laboratory frame; their position enters the detector projection as a linear shift scaled by distance-to-detector. We implement a multi-distance projector: for each of the $D$ detector distances (up to four in typical NF-HEDM configurations), the scattering direction is combined with the grain position and projected onto the detector plane using the MIDAS fifteen-parameter detector geometry (Sharma *et al.*, 2026*a*), yielding fractional $(y, z)$ pixel coordinates and a per-distance validity mask. The full validity mask is the intersection of all per-distance masks (the "AllDistsFound" convention from the C code). For pf-HEDM, the `ScanConfig` adds an outer loop over beam translation positions; a per-scan mask tracks whether each grain is illuminated at each beam position. This block ports the C routine `DisplacementSpots` from `NF_HEDM/src/SharedFuncsFit.c:269-292` and the FF beam-proximity filter from `FF_HEDM/src/FitOrStrainsScanningOMP.c:1048-1058`.

### 4.2.5 Output modes.

The pipeline above terminates in a `SpotDescriptors` dataclass carrying per-spot omega, eta, $2\theta$, $(y, z)$ pixel coordinates, frame number, and validity mask. Two output modes consume this structure:

- `predict_spot_coords` (FF/pf): returns the raw angular and pixel-space coordinates for each valid spot, matching the per-spot output of `ForwardSimulationCompressed`. This is the format consumed by grain-fitting losses that operate directly on spot positions.
- `predict_images` (NF): rasterises each valid spot onto a three-dimensional ($n_{\text{frames}} \times n_y \times n_z$) detector volume via a Gaussian splatting kernel, with configurable sigma and radius. The resulting image is directly comparable with the output of `simulateNF`. Splatting uses soft assignment to the neighbouring pixel grid, preserving gradients with respect to sub-pixel spot positions.

**Intensity model.** Each grain (FF/pf) or voxel (NF) is treated as a point source: peak positions are computed from a single centroid ray, and intensities follow the CIF-based relative-intensity convention used by `ForwardSimulationCompressed` and `simulateNF`, inherited from the same convention in MIDAS (Sharma *et al.*, 2026*a*; Sharma *et al.*, 2026*b*) and HEXRD (Boyce & Bernier, 2013; Bernier *et al.*, 2011). This choice is what enables pixel-exact agreement against the canonical C references (Section 5) and keeps the forward pass differentiable without introducing volume-intersection polyhedral geometry. It also means the model does not natively represent partial illumination when the beam cross-section is smaller than the grain volume, nor the spatially extended peak shapes that the volume-intersection model of `xrd_simulator` (Henningsson & Hall, 2023) captures; in practice, production HEDM indexing pipelines work on spot centroids and this approximation matches what they consume. We return to this point in the Discussion.

### 4.3 Numerical Choices that Preserve Differentiability

Converting a production C forward model into a differentiable one is not a mechanical translation. Several C idioms including hard branching on sign, `acos` at domain boundaries, SVD-based orthogonalisation, integer-rounded pixel coordinates— either break the gradient chain or introduce discontinuities that trap gradient-based optimisers. We made four systematic substitutions:

1. **`torch.where` in place of `if`.** All conditional logic in the pipeline (quadratic vs. degenerate branch, positive vs. negative omega root, per-distance validity) is expressed as masked selection between two fully evaluated paths, not as Python branching. This gives correct gradients on both sides of every decision boundary.

2. **Clamped inverse-trigonometric functions.** All `acos` and `asin` invocations clamp their argument to $[-1+\varepsilon, 1-\varepsilon]$ with $\varepsilon = 10^{-7}$ before evaluation. At the unclamped boundaries the derivative is unbounded; clamping trades an $\mathcal{O}(\varepsilon)$ numerical bias for bounded, finite gradients. In particular, near the azimuthal-axis limit $\eta \to 0, 180^\circ$ and near the reciprocal-space degeneracy $\cos\theta \to \pm 1$, the clamped derivative is finite and approximately correct to order $\sqrt{\varepsilon}$; it does not poison gradient accumulation at the physical boundaries of the forward model.

3. **Newton–Schulz orthogonalisation.** As noted in (1), we avoid SVD-based $SO(3)$ projection because SVD gradients are ill-defined at repeated singular values a situation that arises routinely during optimisation when the orientation matrix drifts.

4. **Validity handling.** Spot validity is enforced by a binary mask (cast to `float32`) that multiplies into all downstream loss terms: invalid spots contribute zero loss and zero gradient, valid spots contribute their full chain-rule gradient through the spot position. The resulting loss surface is piecewise smooth: it is differentiable with respect to all leaf parameters *within* regions where the valid spot set is stable, and exhibits a discontinuity at parameter values for which a spot transitions between the valid and invalid sets. In practice the valid set is large (typically several hundred reflections for a single grain) and its members are well inside the validity region, so gradient-based optimisation rarely traverses a transition during a typical step; empirically (see supplementary SS4) none of the transitions are traversed within the 40-step L-BFGS schedules used below. A continuous (sigmoid-based) relaxation of the validity mask is straightforward to substitute if a future application requires gradient flow across a validity transition.

A further subtle point concerns the detector pixel rounding used in the Gaussian splatter for NF output. A naïve `round` operation has zero gradient almost everywhere; we therefore use a straight-through estimator that exposes gradients from the neighbouring pixel cells back to the sub-pixel spot centre, preserving differentiability of image-space losses with respect to orientation and position.

### 4.4 Gradient Flow

Every input to the forward model is, by construction, a leaf of the PyTorch autograd graph. A scalar loss on the output has well-defined analytic gradients with respect to:

- Grain/voxel orientations (via Euler angles or rotation matrices).
- Grain/voxel positions (in the laboratory frame).
- Per-grain lattice parameters, and hence any strain state expressed in that basis.
- All components of the MIDAS fifteen-parameter detector geometry: sample-to-detector distances $L_{sd}$, beam-centre pixel coordinates $y_{BC}$, $z_{BC}$, pixel size, tilts, and the analytical distortion harmonics.
- Beam parameters: wavelength (which enters the Bragg angle through the B matrix pathway when strain is optimised).
- Beam-translation positions (pf-HEDM).

Parameters that are inherently discrete, the integer Miller indices, the ring assignment, the detector pixel grid—do not participate in the gradient and are treated as fixed buffers. Section 5.6 reports the gradient-correctness check; the per-step-size table is in supplementary SS2.

# 5 Validation

## 5.1 Methodology

For each geometry (FF, NF, pf-HEDM) we construct an identical set of inputs, grain state, detector geometry, omega sweep, reflection list and drive both the C reference simulator and the differentiable PyTorch model with those inputs. Agreement is assessed at two levels: (i) *presence agreement*, the fraction of C-reported spots or pixels that have a corresponding prediction in the PyTorch output; and (ii) *coordinate agreement*, the residual in omega, eta, and $(y, z)$ pixel coordinates for matched spots. The reference hkl list is generated once by `GetHKLList` and shared by both codes, ensuring that discrepancies cannot arise from reflection-set differences.

## 5.2 Far-Field Agreement

A single-grain FF test configuration drawn from the MIDAS regression test suite produces **162 predicted spots in both the C simulator and the differentiable model**, with all 162 matched between the two (100%). Residual errors on the matched spots are:

- omega: mean $2 \times 10^{-4\,\circ}$, maximum $5 \times 10^{-4\,\circ}$
- eta: mean $< 10^{-5\,\circ}$, maximum $10^{-4\,\circ}$
- $(y, z)$ pixel: mean and maximum below $10^{-3}$ px, consistent with floating-point round-off of the pixel-coordinate computation.

At the reported precision the two codes are indistinguishable: the residuals are dominated by the differing order of floating-point operations between a cache-friendly C loop and a vectorised PyTorch einsum, not by any physical disagreement.

## 5.3 Near-Field Agreement

For NF-HEDM we compare synthesised detector volumes produced by `simulateNF` and by the differentiable model's `predict_images` output head over an identical triangular-voxel grid. The C simulator produces 2304 pixel hits across 352 unique spots, and the PyTorch model produces 2304 pixel hits. Frame-by-frame exact-index comparison over all 1,440 $\omega$ frames yields **2304/2304 pixels identical (100.0%)**: the pixel-wise difference map between the two rendered detector integrations is uniformly zero across the entire detector region (Figure 3). The same agreement is preserved when non-zero detector tilts are applied (see `tests/test_tilts.py` for the cross-code validation sweep up to 1° tilt per axis). A standalone end-to-end verification script that runs both pipelines and reports the exact-match count is provided at `paper/scripts/run_side_by_side_nf.py` for reproducibility.

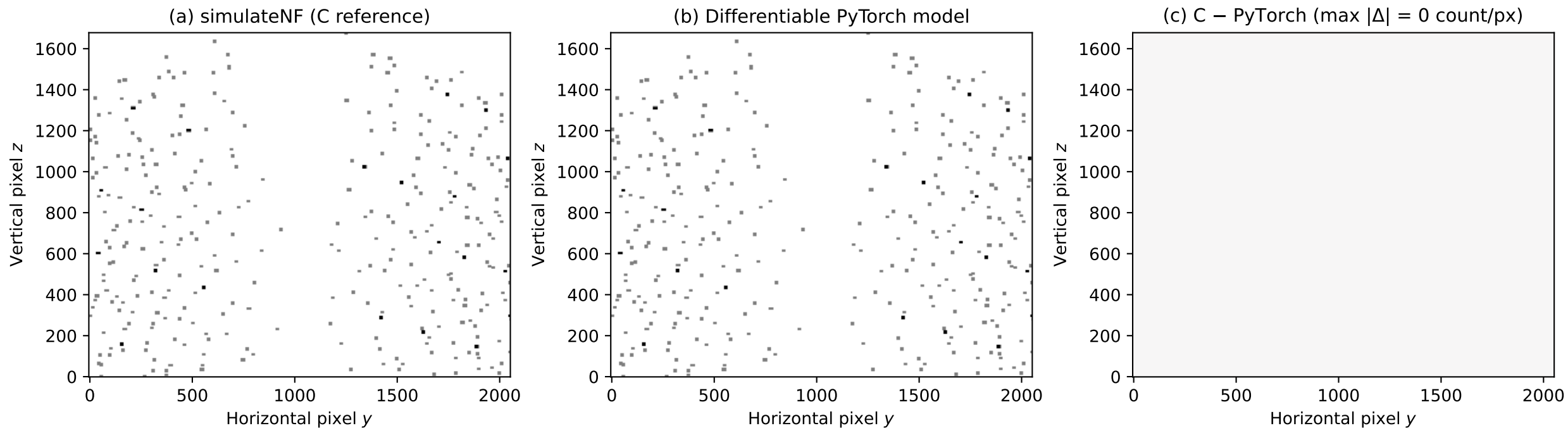

Figure 3: Pixel-exact NF-HEDM cross-code agreement on a single-voxel cubic-Au fixture, summed over all 1,440 $\omega$ frames. (a) `simulateNF` (the C reference). (b) Differentiable PyTorch model. (c) Pixel-wise difference, identically zero everywhere ($\max|\Delta| = 0$ counts/px). The 2,304 pixel hits produced by the C code and the 2,304 produced by the PyTorch model match in $(y, z, \omega)$ index for every reflection.

### 5.4 pf-HEDM Agreement

Point-focused HEDM is tested by running `ForwardSimulationCompressed` in multi-scan mode against `HEDMForwardModel` configured with a matching `ScanConfig`. A five-grain configuration with beam translation across the sample yields **1088/1096 spots matched (99.27%)**; three of the five grains match 100%, while the remaining two each have four unmatched spots. All eight unmatched spots share a common signature: they lie within $0.001\,^\circ$ of $|\eta| = 0^\circ$ or $|\eta| = 180^\circ$, i.e. at the degeneracy point between the quadratic and special-case branches of the omega solver (the $|G_y| <$ `almostzero` boundary in Section 4). In this narrow region, branch selection depends on floating-point comparisons of quantities below the $10^{-12}$ threshold. The two implementations resolve the degeneracy differently: the C code branches via cascaded `if`-statements whose outcomes depend on the exact sequence in which floating-point comparisons are evaluated at compile time, whereas the PyTorch model uses `torch.where` to make a deterministic, vector-consistent branch choice across all reflections (Section 4, numerical substitution (1)). The eight unmatched reflections therefore represent an upper bound on the disagreement zone between two legitimate branch selections near a physical singularity of the forward problem, not a bug in either code; both reproduce the observable diffraction signal to within the physical resolution of the detector. Matched-spot residuals are below $0.03\,^\circ$ in omega and $0.025$ px in pixel coordinates across all five grains, with most grains agreeing below $0.002\,^\circ$ and $0.002$ px.

### 5.5 Summary Table

Table 1 summarises the three geometries. The physical-resolution takeaway is consistent across modes: the differentiable PyTorch model reproduces the canonical C simulators to within floating-point precision everywhere except at well-understood numerical-threshold boundaries.

Table 1: Cross-code agreement between the differentiable PyTorch forward model and the canonical C reference simulators. Residuals are reported for matched spots/pixels only.

| Mode | Reflections | Match rate | Max $\omega$ err | Max pixel err |
|---|---|---|---|---|
| FF | 162/162 | 100.0% | $5 \times 10^{-4}$ ° | $< 10^{-3}$ px |
| NF | 2304/2304 | 100.0% | 0 | 0 px (indexed) |
| NF (with tilts) | 2322/2322 | 100.0% | 0 | 0 px (sweep up to 1°) |
| pf-HEDM | 1088/1096 | 99.27% | 0.03 ° (edge cases) | 0.025 px |

### 5.6 Gradient Correctness

Every parameter of the forward model is a leaf node of the autograd graph. We verified analytic-vs-finite-difference agreement on a single-grain FF-HEDM configuration (cubic Au, 162 valid spots), with the loss equal to the sum of squared $\omega$, $y$, and $z$ residuals over all valid spots, sweeping the relative finite-difference step from $10^{-3}$ to $10^{-6}$. The detector-geometry parameters ($L_{sd}$, $y_{\mathrm{BC}}$, $z_{\mathrm{BC}}$) agree with autograd to $10^{-8}$ relative error at every step size tested, indicating that the loss surface is locally well-approximated by its tangent and the residual is dominated by floating-point round-off. Grain-position gradients agree to $10^{-9}$ at the coarsest step and degrade to $\sim 10^{-5}$ at the finest as cancellation noise grows. The Euler-angle gradient agrees to $7.8 \times 10^{-8}$ at the intermediate $10^{-5}$ step but is degraded at the coarsest step: this is the soft-mask boundary signature described in Section 4 – a coarse perturbation moves some predicted spots across the eta-validity cutoff, and finite-difference correctly picks up a step contribution that autograd correctly does not. Restricting finite differences to step sizes small enough to stay inside the validity cells restores tight agreement. The full per-step-size table is given in supplementary SS2; the optimisation range exercised in Section 6 shows no elevated residuals.

### 5.7 At-Scale and Noisy Cross-Code Agreement

A 250-grain synthetic FF-HEDM population ($\sim$ 41,000 reflections) cross-validates against the reference simulator with the reference's `SimNoiseSigma` parameter set to 0, 0.5 px, and 2.0 px. The match rate is 99.4–99.6% across all three regimes, the shortfall is the near-axis branch-boundary set described above, and the noisy-regime residuals match the analytic Rayleigh prediction for a two-component Gaussian shift to within 1% on the mean and $\sim$ 8% on the extreme. Full numbers and the radial-residual overlay are given in supplementary SS3.

## 6 Gradient-Based Optimisation

### 6.1 Basin of Convergence

A practical user of this framework needs to know how close an initial guess must be for gradient-based refinement to succeed. We quantify this on a single FF-HEDM grain (cubic Au, 162 valid spots) by sweeping the initial orientation perturbation across ten levels from 0.5° to 30°, with $N = 20$ trials per level (uniformly-sampled rotation axis, independent lattice perturbations), and recording the fraction of trials that converge to within $0.1°$ of the truth after a three-phase L-BFGS schedule (orientation, then strain, then jointly).

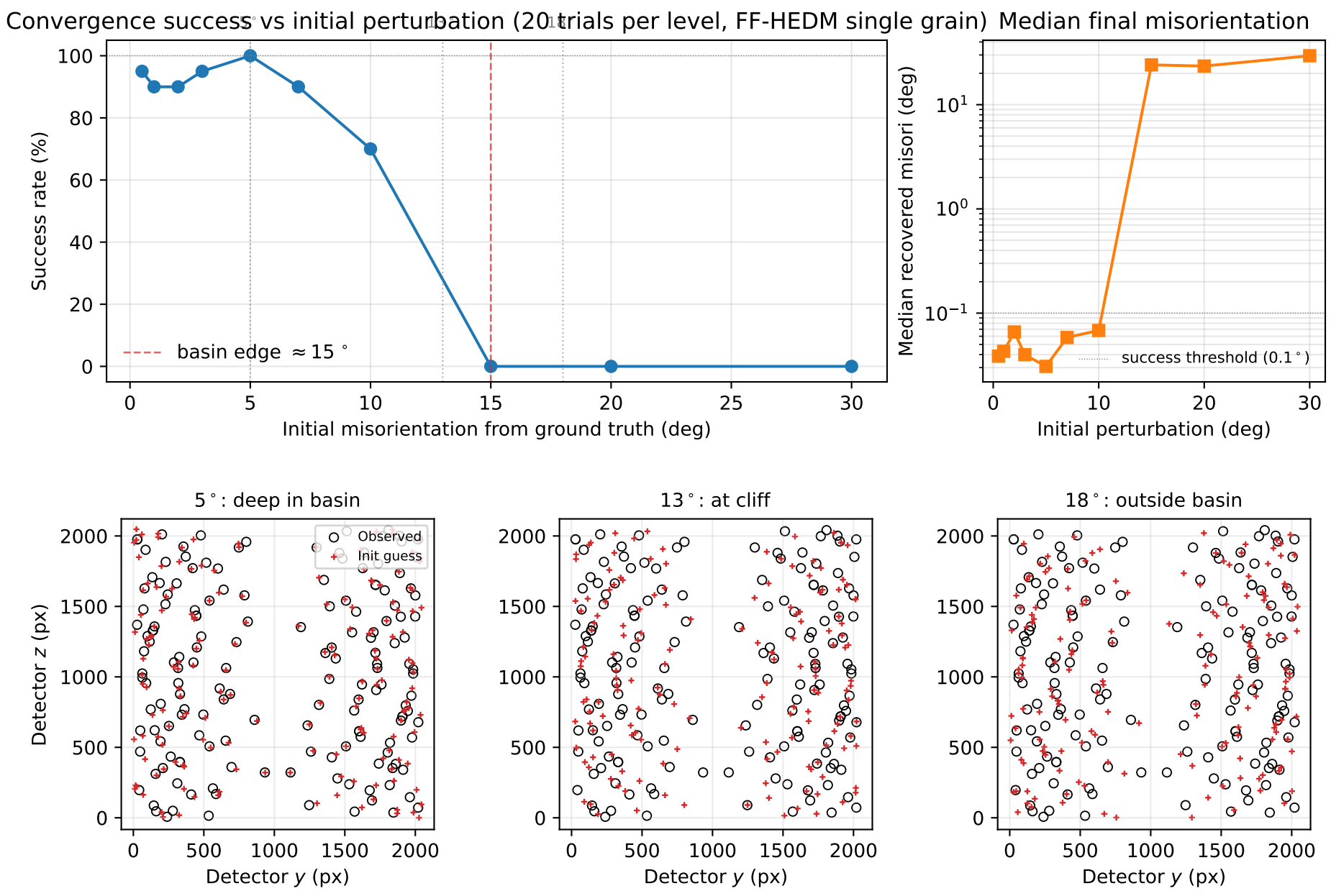

Figure 4: Basin of convergence for three-phase L-BFGS joint refinement on a single FF-HEDM grain. Success fraction (recovered misorientation $< 0.1°$) is $\geq 90\%$ for initial perturbations up to $10°$, with a sharp cliff at $\sim$ 12–15°. Full sweep details and representative spot-overlay plots in supplementary SS5.

Figure 4 summarises the result: $\geq 90\%$ recovery to $0.1°$ for initial perturbations up to $10°$, with a sharp cliff near 12–15° where the nearest-neighbour spot-matching assignment inside the loss closure begins to misroute predictions to wrong observations. For practical use this is reassuring: production FF-HEDM indexers (Sharma *et al.*, 2026*a*) typically deliver initial orientations within $5°$, comfortably inside the basin. The full sweep methodology, all twenty success-rate points, and representative spot-overlay panels at perturbations spanning the cliff are given in supplementary SS5.

## 6.2 A Note on Grain Position in Far-Field Geometries

The original scope of this demonstration included joint recovery of grain position together with orientation and strain. In the far-field geometry used above ($L_{sd} = 1\,\text{m}$), an 8-µm grain position offset in any laboratory axis produces only a $\sim$ 0.04 pixel shift in the detector coordinates of the predicted spots—well below both the pixel-quantisation threshold and the measurement noise floor. Position therefore enters the FF-HEDM forward model through a nearly-singular sensitivity Jacobian, and its recovery at micrometer-level accuracy is physically unconstrained by FF data alone. This is not a property of the differentiable framework but of the geometry: the same degeneracy exists in any FF-HEDM reconstruction method.

Grain position becomes directly recoverable in near-field HEDM (where small sample-to-detector distances magnify the position-to-pixel response) and in pf-HEDM (where beam translation independently constrains position along the beam axis). Demonstration of gradient-based position recovery in those geometries is a natural follow-on.

## 6.3 NF-HEDM Joint Recovery of Orientation, Strain, and Position

Because grain position couples weakly to FF detector coordinates at $L_{sd} = 1\,\mathrm{m}$ (Section 6.2), we demonstrate the framework's ability to *jointly* optimise orientation, strain, and position by running the same pipeline in near-field geometry, where $L_{sd} = 5\,\mathrm{mm}$ and $\mathrm{px} = 1.48\,\mu\mathrm{m}$ make position directly observable: a $1\,\mu\mathrm{m}$ shift in any laboratory axis produces a $\sim 0.68\,\mathrm{px}$ shift in the diffracted-ray detector coordinates. This is the regime in which NF-HEDM reconstructs grain shape to begin with, and it is the regime in which the differentiable framework unlocks a capability that the existing C-based HEDM pipelines cannot straightforwardly replicate: joint *continuous* optimisation across all three parameter classes in a single gradient-flow.

**Setup.** A single cubic-Au voxel at ground-truth position $(15, -8, 5)\,\mu\mathrm{m}$ with the same tetragonally-distorted lattice and Euler angles as Section 6. Initial guess: orientation perturbed by $1.7°$, lattice edges by up to $5 \times 10^{-3}\,\text{Å}$, and position by up to $\sim 16\,\mu\mathrm{m}$ (roughly 10 pixels in the detector projection).

**Optimisation schedule.** Four L-BFGS phases:

1. Orientation only, angular-coordinate loss (position-independent).
2. Strain only, angular-coordinate loss.
3. Position only, detector-coordinate loss.
4. Joint refinement of orientation, strain, and position, detector-coordinate loss.

Angular loss is used in the first two phases because it is invariant to grain position, letting orientation and strain converge independently of the poor initial position estimate. Detector-space loss is used in the position and joint phases because position enters the forward model only through the detector projection.

**Results.** Table 2 and Figure 5 report convergence in both noise-free and noisy modes. The noise-free run recovers position to $\sim 6\,\mathrm{nm}$, orientation to $\sim 10^{-4\,\circ}$, and lattice edges to $\sim 4 \times 10^{-6}\,\text{Å}$. Under resolution-matched Gaussian noise, position recovers to $\sim 66\,\mathrm{nm}$, orientation to $\sim 5 \times 10^{-3\,\circ}$, and lattice to $\sim 3 \times 10^{-4}\,\text{Å}$. The decoupled schedule—angular-loss phases first, then detector-loss phases—is essential: starting Phase 3 directly on a joint detector-space loss would have all three parameter classes competing to explain pixel residuals whose dominant contribution is position, masking the orientation and lattice channels.

Table 2: Single-grain NF-HEDM joint recovery of orientation, strain, and grain position, starting from a 1.7° misoriented, $5 \times 10^{-3}$ Å lattice-perturbed, 16 $\mu$m position- perturbed initial guess. Near-field geometry with $L_{sd} = 5$ mm and px $= 1.48\,\mu$m.

| Mode | Final misorientation | Max lattice-edge error | Max position error |
|---|---|---|---|
| Noise-free | $5.6 \times 10^{-5}$ ° | $4.2 \times 10^{-6}$ Å | 0.006 $\mu$m |
| Noisy | $5.4 \times 10^{-3}$ ° | $2.7 \times 10^{-4}$ Å | 0.066 $\mu$m |

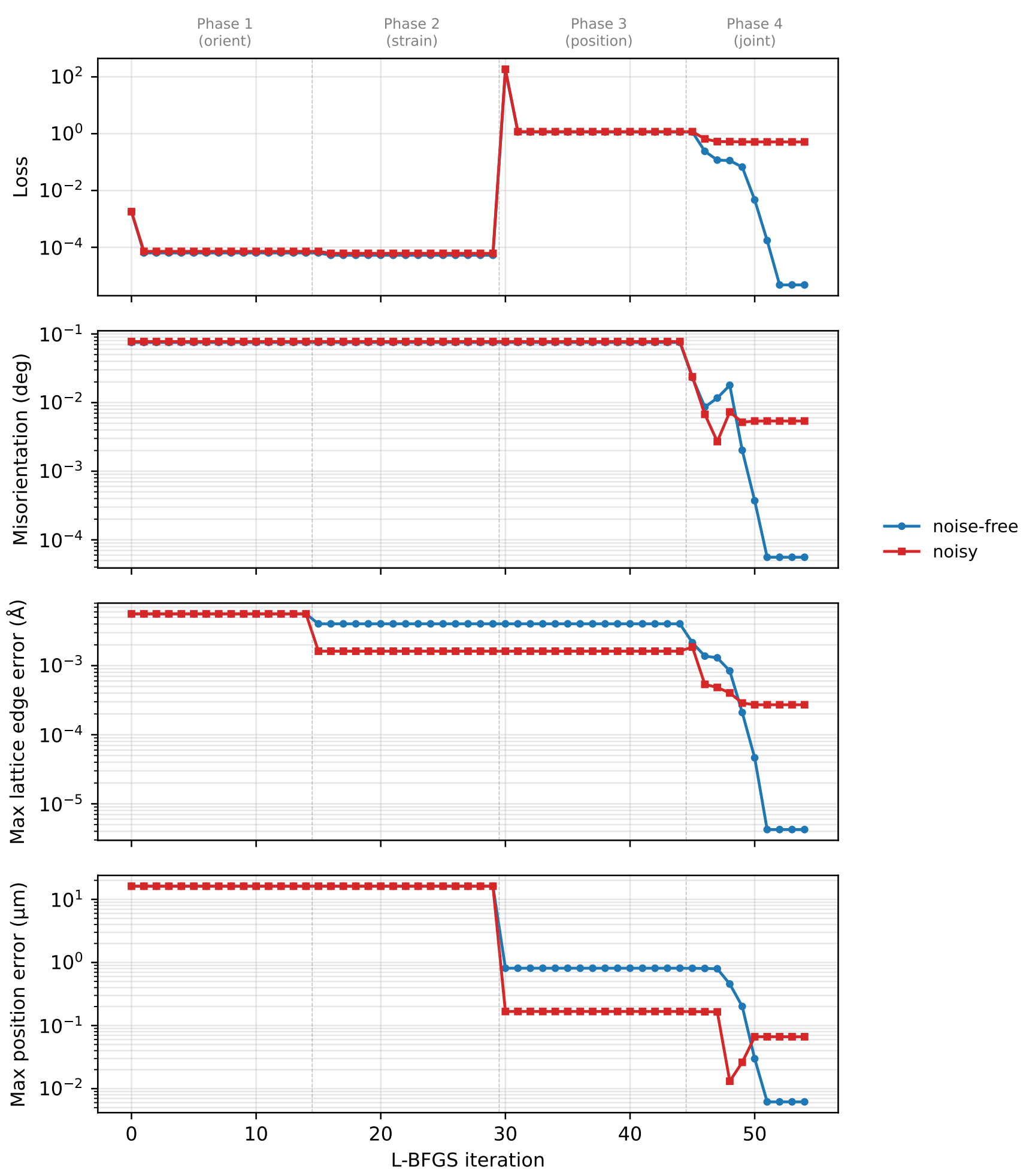

Figure 5: NF-HEDM four-phase L-BFGS joint recovery of orientation, strain, and grain position from synthetic spots. The initial guess is 1.7° misoriented, with lattice parameters perturbed by $\sim 5 \times 10^{-3}$ Å and grain position perturbed by up to 16 $\mu$m. Phases 1–2 bring orientation and strain to sub-0.1° and $10^{-4}$ Å via angular-coordinate losses; Phase 3 closes the position gap with a detector-space loss; Phase 4 jointly polishes all three. Final position error: 6 nm noise-free, 66 nm under resolution-matched noise. Dashed vertical lines separate the four phases.

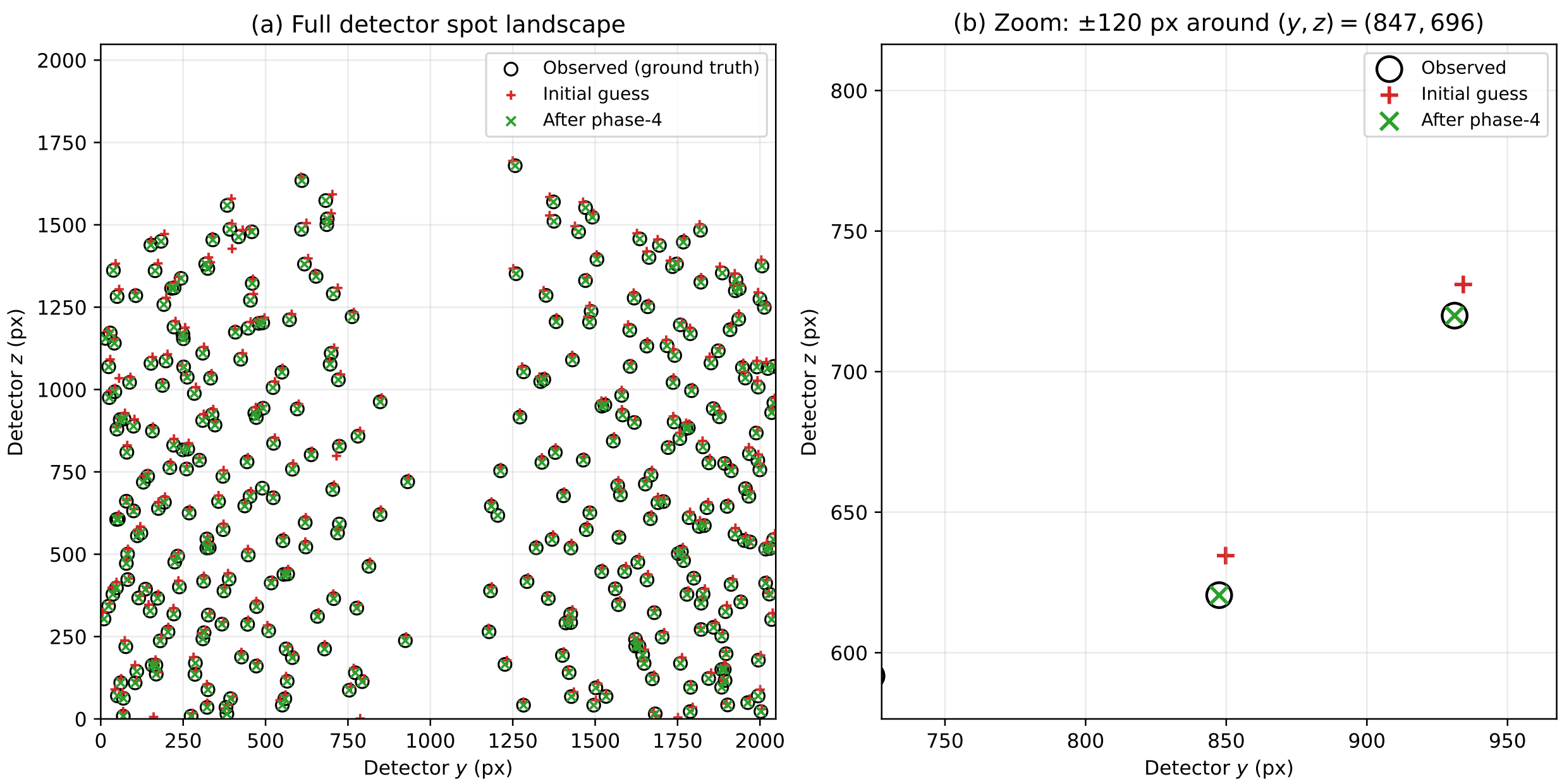

Figure 6: Spot-level recovery for the same NF-HEDM joint optimisation as Figure 5. Black open circles: observed (ground-truth) detector positions. Red "+" markers: predicted positions at the perturbed initial guess (1.7°, $\sim 16\,\mu$m). Green "×" markers: predicted positions after the four-phase L-BFGS recovery. (a) Full detector view — after recovery, the green "×" markers sit on the black observed circles for every spot; the red initial guesses are visibly offset. (b) Representative zoom showing typical residual offset of the initial guess and pixel-exact reduction of that offset after Phase 4.

**Relation to existing workflows.** Production NF-HEDM reconstruction pipelines (Suter *et al.*, 2006; Sharma *et al.*, 2026*a*) solve for per-voxel orientations on a fixed geometric grid; grain position is not an optimisation variable but an output of the reconstructed voxel assignments. The differentiable framework additionally permits *continuous* refinement of sub-voxel position as a simultaneous optimisation variable alongside orientation and strain. This is the kind of capability that existing gradient-free pipelines cannot straightforwardly deliver—the parameter space has $7 + 3 = 10$ continuous degrees of freedom per grain, and joint coordinate-descent on each is expensive without gradients. The demonstration here closes that gap.

### 6.4 Experimental Validation on Real FF-HEDM Data

As a final validation we apply the differentiable forward model to a real FF-HEDM reconstruction: an $\alpha$-titanium alloy sample (space group 194, hexagonal, $a = 2.927$ Å, $c = 4.675$ Å) measured at the 1-ID-E beamline (Advanced Photon Source) with $L_{sd} = 940.7$ mm, $\lambda = 0.189714$ Å, and a $2048 \times 2048$ detector at $200\,\mu$m pixel pitch over a 720° omega sweep (0.25° step). The MIDAS production pipeline reconstructed 214 grains with 48,164 measured spots in `SpotMatrix.csv`. We perform three independent experiments against this reconstruction: a forward-model agreement check (which tests our physics against MIDAS's own refined fit), a perturb-and-recover experiment (which tests our optimiser on real measurement-noise correlations), and a basin-of-convergence sweep (which measures the practical basin on real data).

**A note on what `SpotMatrix.csv` contains.** MIDAS's `SpotMatrix.csv` stores measured spot centroids after the production DetCor step (corrected for detector tilts and the 15-parameter empirical distortion model, from the `YOrig(DetCor)/ZOrig(DetCor)` columns of the internal `InputAllExtraInfoFittingAll.csv` file). Our FF forward model applies a tilt-free projection precisely because of this: the measurements are already geometry-corrected, and applying tilts in the model would double-correct. The NF framework is separately validated against `simulateNF` with non-zero tilts at 100% pixel agreement up to 1° per axis (`tests/test_tilts.py`).

**Test 1: Per-grain head-to-head against MIDAS's production fit (Figure 7b).** For each of the 214 grains we run a single-shot 12-degree-of-freedom joint L-BFGS optimisation (3 position, 3 orientation, 6 strain) over the real measured spots and compute per-grain aggregate residuals in the same form as MIDAS's `DiffOme`, `DiffAngle`, and `DiffPos` columns. The comparison is between two methodologically distinct optimisers on identical data. MIDAS's production fit decomposes the per-grain refinement into a three-step sequential coordinate-descent—position-only fit (3 dof), then orientation-only fit (3 dof), then strain fit (6 dof), each step holding the others fixed and using finite-difference Jacobians. Our model jointly refines all twelve parameters in a single L-BFGS run with autograd-supplied exact gradients, exploiting cross-parameter coupling that the decoupled schedule discards. The two procedures are therefore not expected to lie on the $y = x$ diagonal of panel (b) by construction; what matters is the asymmetry of the scatter about it. The favourable region is *below* the diagonal, where the joint optimiser delivers a smaller per-grain residual than the sequential fit.

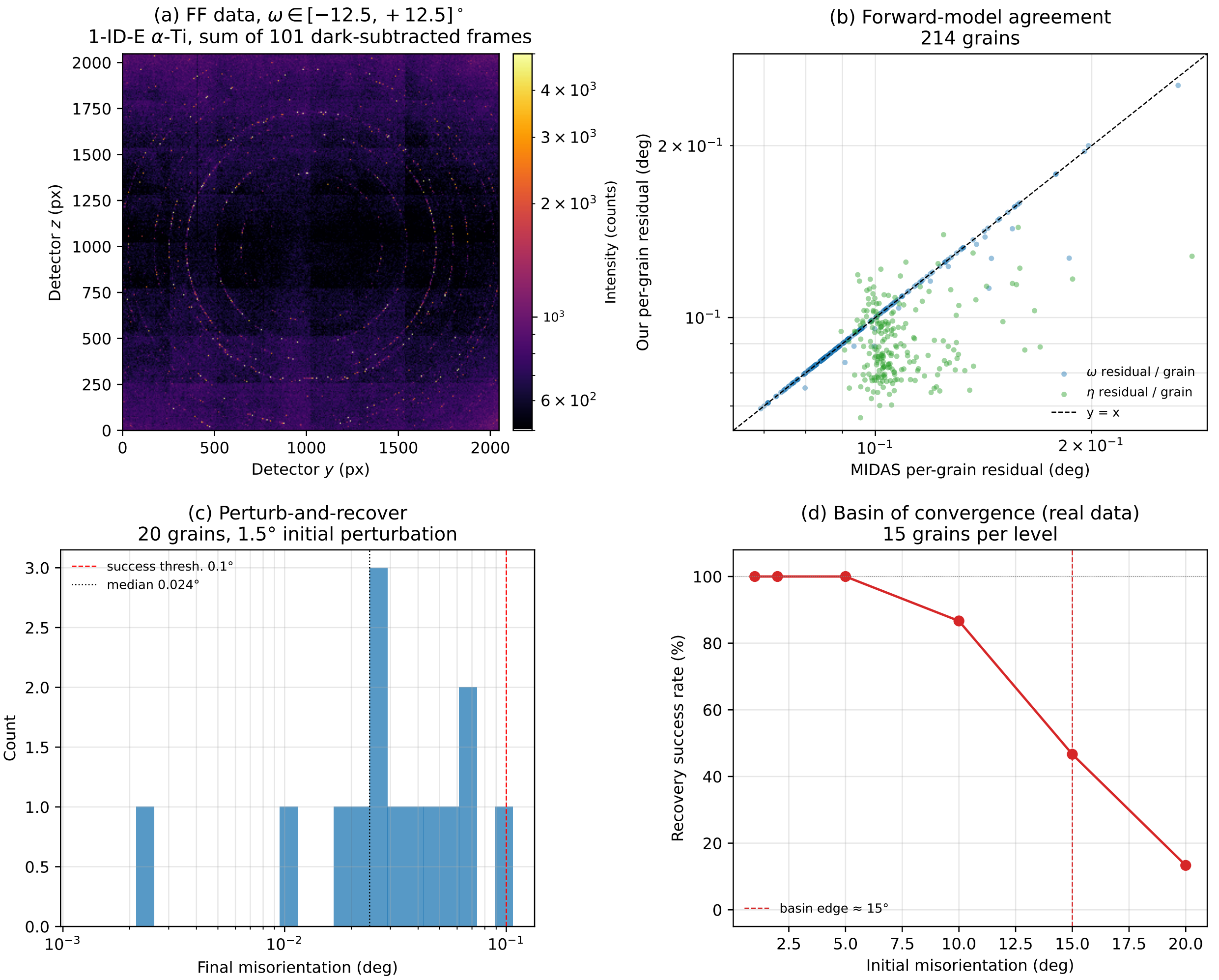

Figure 7: Experimental validation on the 214-grain $\alpha$-Ti FF-HEDM reconstruction at 1-ID-E. (a) Sum of 101 dark-subtracted FF detector frames around $\omega = 0$ ($\omega \in [-12.5^\circ, +12.5^\circ]$ at $0.25^\circ$ per frame), shown on a log intensity scale. Single FF frames are too sparse to be visually informative in print; the small rotation sum reveals both the Debye–Scherrer powder rings and the individual indexed Bragg spots that the next three tests are run against. (b) Per-grain residuals of our 12-dof joint L-BFGS fit (single shot, autograd gradients) versus MIDAS's three-step sequential position/orientation/strain fit on the same 214 grains. The two methods optimise different objectives, so the $y = x$ diagonal is a reference, not an equality target; below-diagonal indicates the joint fit delivers a smaller per-grain residual than the sequential fit. Our model is below the diagonal on 179/214 grains (83.6%) for both $\eta$ and position and 120/214 (56.1%) for $\omega$; medians are tied on $\omega$ and 14.7%–39% tighter on $\eta$ and position. (c) Histogram of final misorientation after three-phase L-BFGS recovery on 20 grains perturbed by 1.5° about a random axis: 100% recovered to below 0.1°, median 0.024°. (d) Real-data basin of convergence: recovery success rate against initial misorientation from 1° to 20°, 15 grains per level. Basin edge sits sharply at 10–12°.

Table 3: Real-data basin of convergence for three-phase L-BFGS recovery on the $\alpha$-Ti FF-HEDM dataset (15 grains per perturbation level, success threshold $< 0.1°$).

| Initial misori (deg) | Success rate | Median final (deg) |
|---|---|---|
| 1.0 | 15/15 (100%) | 0.026 |
| 2.0 | 15/15 (100%) | 0.042 |
| 5.0 | 15/15 (100%) | 0.024 |
| 10.0 | 13/15 (87%) | 0.035 |
| 15.0 | 7/15 (47%) | — |
| 20.0 | 2/15 (13%) | — |

The aggregate medians and the below-diagonal counts fall on that side. Median absolute $\omega$ residuals are essentially tied (0.0901° vs 0.0904°), with 120/214 grains (56.1%) below the diagonal. Median $\eta$ residuals are 0.0879° vs 0.1031° (joint fit 14.7% tighter), with 179/214 grains (83.6%) below the diagonal. Median per-grain position residuals are 134 $\mu$m vs 218 $\mu$m (joint fit 39% tighter), again with 179/214 grains (83.6%) below the diagonal. The joint optimiser thus matches MIDAS on $\omega$ and is materially tighter on $\eta$ and position—the expected signature of exact gradients through coupled parameters extracting information that a decoupled three-step descent cannot. If our forward model were missing physics, the residuals at the MIDAS state would scatter above the diagonal, not below.

**Test 2: Perturb-and-recover on real measured spots (Figure 7c).** We take the MIDAS-reconstructed state of the 20 highest-confidence grains, perturb the orientation by 1.5° about a random axis and perturb the lattice parameters by $\|\Delta \mathbf{l}\| \leq 2 \times 10^{-3}$ Å, and run our three-phase L-BFGS schedule against the real measured spots from `SpotMatrix.csv` (not simulated observations). All 20 grains recover to below the 0.1° success threshold; the median final misorientation is 0.024°, maximum 0.089°. Median final lattice-edge error is $1.8 \times 10^{-3}$ Å. This is the critical test of the optimiser against real measurement-noise correlations rather than the synthetic Gaussian noise of Section 6: the same three-phase schedule that converges on synthetic data also converges on measurement-noise data, at the same precision.

**Test 3: Real-data basin of convergence (Figure 7d).** We sweep the initial perturbation from 1° to 20° (steps: $1, 2, 5, 10, 15, 20°$) on 15 grains per level. Results:

The basin edge on real data sits at approximately 10–12°, very close to the 12–15° edge measured on synthetic data (Section 6.1). The small tightening reflects real measurement noise contributing to the loss-surface structure, but production FF-HEDM indexers deliver initial orientations comfortably inside this basin ($< 5°$ is typical (Sharma *et al.*, 2026*a*)), so the result is operationally relevant: our optimiser, handed any production-quality indexer output, converges cleanly against real measured spots.

**Summary.** On real 1-ID-E data, the differentiable forward model (i) delivers per-grain residuals at parity with or below MIDAS's three-step sequential refined fit (median $\omega$ tied, median $\eta$ 14.7% tighter, median position 39% tighter; below-diagonal on 83.6% of grains for $\eta$ and position), (ii) recovers the reference state from 1.5° perturbations with 100% success on measurement-noise observations, and (iii) has a real-data basin of convergence of $\sim 10°$ that covers the typical precision of production indexers. The framework's optimisation machinery is not merely a synthetic demonstration — it works on the same data used by the production MIDAS pipeline.

**Differentiable auto-calibration.** Because the tilt parameters enter the forward model through an autograd-compatible rotation composition, they are natively optimisable. Setting `model.tilts.requires_grad = True` and including `model.tilts` in any gradient-based optimiser (`Adam`, `L-BFGS`, etc.) makes the 3-parameter tilt state directly learnable. The gap this closes is specific to NF-HEDM. FF and pf-HEDM produce a powder background alongside the indexed-grain spots, and the MIDAS `AutoCalibrateZarr` and `CalibrantPanelShiftsOMP` routines refine the full detector geometry (including tilts, beam centre, and distortion harmonics) from that powder signal to sub-pixel precision before indexing begins; the same capability is provided by pyFAI for azimuthal integration (Sharma *et al.*, 2026*a*). NF-HEDM, by contrast, produces *only* the indexed-grain spot pattern — there is no powder ring on the NF detector — so none of these powder-based calibration tools apply. In production workflows NF detector geometry is therefore aligned once at experiment start and re-used across the campaign; any drift that accumulates between alignments is undetected and propagates into the reconstruction. The capability demonstrated here is, to our knowledge, the first automated NF-HEDM detector-calibration tool: it closes the loop from indexed-grain observations back to detector geometry in a single gradient flow, without requiring a powder standard, and it is composable with the full HEDM reconstruction pipeline so that drift checks can be run in-line with routine analysis. Supplementary SS6 reports a worked example: starting from a zero-tilt initial guess, the solver recovers a $(0.50°, -0.30°, 0.20°)$ truth tilt to $\sim 0.1\,\mu$rad RMS in 0.21 s on a single CPU core.

## 6.5 Multi-Panel FF-HEDM Joint Geometry-and-Grain Refinement

Modern far-field experiments increasingly rely on multi-panel detectors arranged around the X-ray beam, with the panels' inner corners surrounding a hole through which the direct beam passes. Each panel carries its own distance, beam-centre, and tilt; the detector cluster as a whole is typically rotated by a small angle (10–20°) about the beam axis so that diffraction spots avoid the inter-panel gaps along the cardinal directions. Powder calibration with a standard sample constrains the sample-to-panel distance, the beam-centre, and the in-plane tilts $t_y, t_z$ to high precision but cannot constrain the rotation $t_x$ of each panel about the beam axis – powder rings are invariant under that rotation, by construction. Single-crystal data is therefore required to fix $t_x$, and conventional FF-HEDM pipelines refine panels one at a time in a sequence of independent fits. We use the differentiable forward model to refine all panel geometries *and* all grain states jointly, in one gradient flow, on a synthetic four-panel demonstration dataset.

**Setup.** Four $2048 \times 2048$ panels with $200\,\mu$m pixels are arranged at $L_{sd} \approx 1$ m and nominal panel $t_x$ values $\{15.0°, 105.0°, -75.0°, -165.0°\}$. Each panel's beam centre sits 52 pixels off-detector at the inner corner ($y_{\mathrm{BC}} = z_{\mathrm{BC}} = 2100$ px), so each panel captures one azimuthal wedge of every diffraction cone whose $2\theta$ is large enough to reach it. We populate the volume with ten randomly-oriented Au grains carrying a small deviatoric strain ($\|\varepsilon_{\mathrm{dev}}\| \leq 5 \times 10^{-4}$) and a $\pm 100\,\mu$m position spread, simulate 5,760 diffraction spots through the `multi_mode="panel"` forward (Section 4, with panel-routing and per-detector tilts; reproduced in `scripts/make_4det_dataset.py`), and add Gaussian $\sigma = 0.5$ px noise to every centroid. Figure 8(a) shows the panel layout and the resulting spot population.

**Refinement.** Two scenarios reflect realistic calibration states. **Scenario A** (*tx-only*) starts from the truth geometry except that each panel's $t_x$ is perturbed by a uniform $\pm 0.2°$ – the residual ambiguity left by powder calibration. **Scenario B** (*full setup drift*) additionally perturbs each panel's $L_{sd}$ by $\pm 2$ mm, in-plane tilts $t_y, t_z$ by $\pm 0.1°$, and beam-centre by $\pm 2$ px, modelling a campaign-scale calibration drift. In both scenarios grains are initialised with a small additive Euler perturbation ($\sigma = 0.05°$) and a $\sigma = 5\,\mu$m position offset to emulate a lightly imperfect indexer output, and we refine all of panel-$t_x$ (Scenario A) or all of $\{L_{sd}, y_{\mathrm{BC}}, z_{\mathrm{BC}}, t_x, t_y, t_z\}_d^4$ (Scenario B), together with the per-grain orientation, position, and strain, in a single shared gradient flow. The loss is the elementwise minimum over the two Friedel partners of the squared per-spot residual in $(\Delta y, \Delta z, \Delta\omega/\Delta\omega_{\mathrm{step}})$; the `min`-of-Friedels avoids a fixed K-assignment between the two sign-of-$\omega$ branches and stays smooth as grains rotate during optimisation.

**Why a naive joint L-BFGS gets stuck and how Adam fixes it.** A direct L-BFGS pass over the joint parameter set converges prematurely. Inspecting the gradient at initialisation reveals why: $\|\nabla_{\mathbf{e}}\mathcal{L}\| \approx 2.4 \times 10^3$ for the ten grain Euler triples, while $\|\nabla_{\mathbf{t_x}}\mathcal{L}\| \approx 50$ for the four panel $t_x$ values – an almost $50\times$ ratio. Strong-Wolfe L-BFGS' first step moves in the gradient-norm-dominant direction, the grains absorb the panel tilt error, and the residual loss falls into a non-truth local minimum that the Hessian approximation cannot escape. Across 10 seeded perturbations per scenario, naive joint L-BFGS reaches the noise floor on 3/10 runs in Scenario A and 0/10 runs in Scenario B (Supplementary Fig. S4a); the panel-$t_x$ error after "refinement" is essentially the input perturbation.

We resolve this with a short Adam warm-up (per-parameter learning rates: tilts $5 \times 10^{-2}$, $L_{sd}$ $5 \times 10^1$ on top of a $\times 10^3$ gradient hook to internally use millimetres, $y/z_{\mathrm{BC}}$ 0.1, Eulers $10^{-4}$, positions 1.0, strains $10^{-5}$) of 100 steps before handing over to L-BFGS. Adam's per-parameter normalisation gives panel-$t_x$ enough early progress that L-BFGS' subsequent line search no longer sees the grain compensation as the cheapest descent direction. With the warm-up the joint schedule reaches the noise floor on 25/25 seeds in both scenarios, and the panel-$t_x$ error collapses by another two orders of magnitude (Supplementary Fig. S4a). The per-variant loss trajectories are reported in Supplementary Fig. S4b.

(b)

(c)

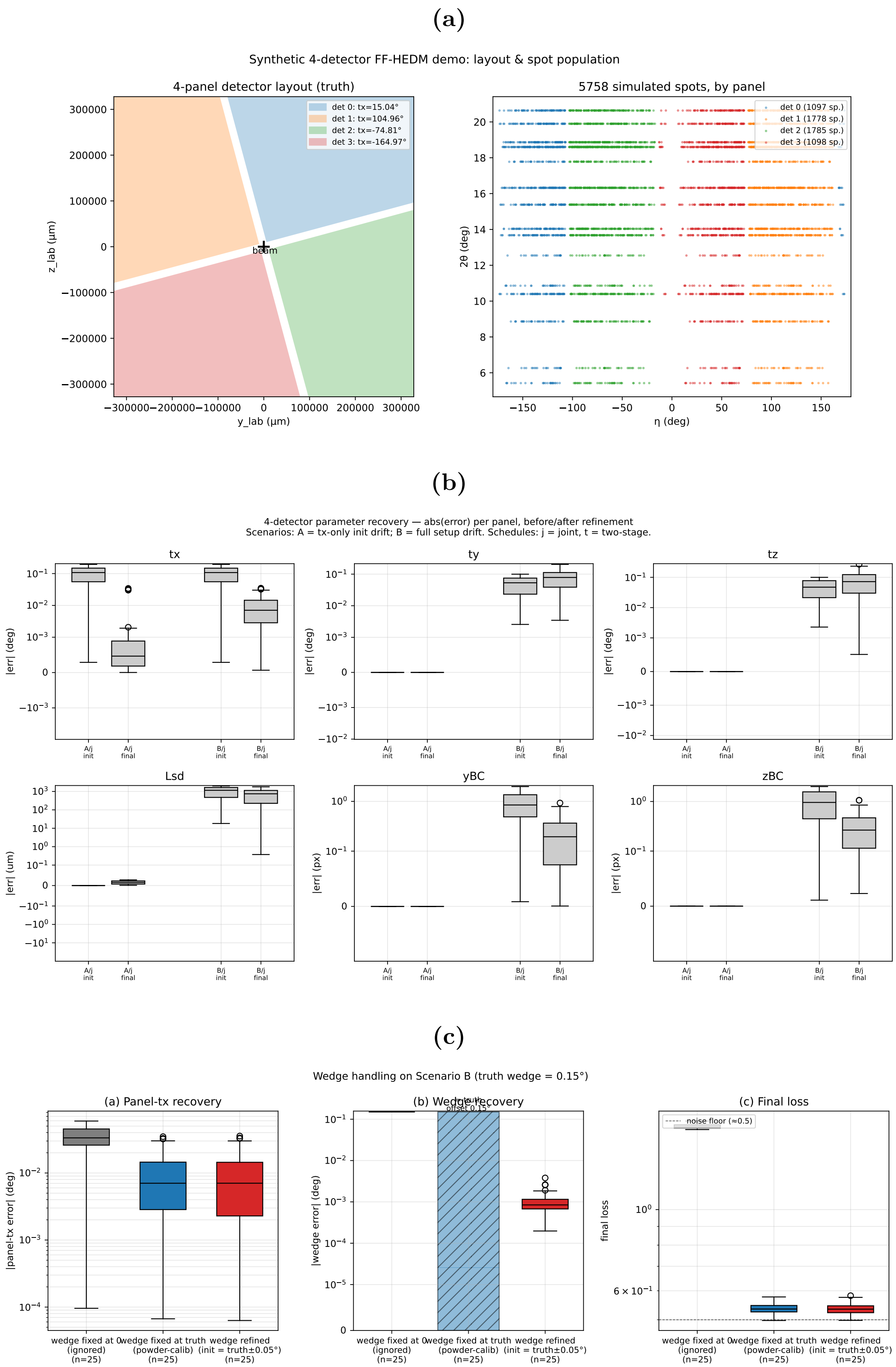

Figure 8: Synthetic four-panel FF-HEDM joint refinement (full detail in text). **(a)** Panel layout (left) and 5,760 forward-simulated spots coloured by panel (right); the assembly is rotated about the beam axis with per-panel $t_x \in \{15, 105, -75, -165\}^\circ$. **(b)** Per-parameter recovery: absolute error before/after refinement on a symmetric-log axis (25 seeds per scenario; “j” = joint Adam-then-L-BFGS, “t” = two-stage). Panel $t_x$ collapses by more than two orders of magnitude; $L_{sd}$ in Scenario B shows the noise-floor data-likelihood degeneracy discussed in the text. **(c)** Wedge

**Results.** Across 25 seeded perturbations per scenario, the joint Adam-then-L-BFGS schedule reaches the per-spot noise floor on 50/50 runs. Panel-$t_x$ recovery is excellent in both scenarios: RMS error across all four panels and all 25 seeds is $7.6 \times 10^{-4\,\circ}$ (13 $\mu$rad) in Scenario A and $1.2 \times 10^{-2\,\circ}$ (210 $\mu$rad) in Scenario B, both more than an order of magnitude tighter than the $\pm 0.2^\circ$ initial perturbation. Mean panel-$t_x$ residual in Scenario A is $5 \times 10^{-4\,\circ}$ ($\sim 9\,\mu$rad), more than two orders of magnitude below the input drift. The two-stage L-BFGS schedule of the original plan is also reported in Figure 8(b) for comparison; it reaches the noise floor on 20/25 Scenario A seeds and only 9/25 Scenario B seeds, trailing joint Adam-then-L-BFGS by roughly 30× on the panel-$t_x$ RMS in either scenario.

Figure 8(b) reports the full per-parameter recovery distribution. In Scenario B the recovery of $L_{sd}$, $t_y, t_z$ and beam-centre is partial: at the noise floor multiple parameter combinations are observationally indistinguishable within the 0.5 px per-spot uncertainty, and the optimiser settles on one of these statistically-equivalent solutions rather than the truth. This is a feature of the data-likelihood and not of the differentiable framework – the same indeterminacy would manifest in any gradient-based or non-gradient-based fit at this noise level. The panel-$t_x$ value is the exception: its action on the prediction (a panel-specific azimuthal rotation of every spot on that panel) is structurally distinct from any combination of the other panel-level or grain-level parameters, and it is recovered cleanly even when the others are not.

**Stress test.** A scenario C (5×-harder perturbation: $\pm 1^\circ$ panel $t_x$, $\pm 0.5^\circ$ in $t_y, t_z$, $\pm 5$ mm in $L_{sd}$, $\pm 5$ px in beam-centre) still converges: from an initial loss of $\sim 6 \times 10^3$ the same joint schedule reaches loss $\approx 1.0$, with a panel-$t_x$ residual of $0.07^\circ$ RMS, a 14× reduction. The capability is therefore not bound to a narrow "almost-already-calibrated" regime: it is robust to realistically large initial calibration errors.

**Wedge.** The wedge angle, defined as the non-orthogonality of the rotation axis and the incident beam, is a single global geometric scalar shared across all panels. We implement it as the rigorous tilt of the rotation axis, $\hat{n} = (\sin W, 0, \cos W)$, so that the rotation about $\hat{n}$ factors as $R_{\hat{n}}(\omega) = R_y(W)\, R_z(\omega)\, R_y(-W)$; the standard omega quadratic then operates on $G' = R_y(-W)\, G_{\text{sample}}$ with the substitutions $G_{x,\text{eff}} = \cos W \cdot G'_x$, $G_{y,\text{eff}} = \cos W \cdot G'_y$, $v_{\text{eff}} = \sin\theta\, |G| + \sin W \cdot G'_z$, and the existing branch-selection logic carries through unchanged. At $W = 0$ this collapses bit-identically to the no-wedge solver. For the four-panel synthetic demonstration we set the truth wedge to $W = 0.15^\circ$ – a value within the residual indeterminacy a powder calibrant leaves behind – and run Scenario B in three regimes: *ignore* (the refinement assumes $W = 0$, mimicking a workflow that omits the wedge correction); *fixed-truth* (the refinement is given the truth wedge from the powder calibrant, the realistic baseline); and *refine* (the refinement starts at $W_{\text{truth}} \pm 0.05^\circ$ and opts wedge into the joint optimiser). Figure 8(c) reports the recovery distribution. Ignoring the $0.15^\circ$ wedge inflates the panel-$t_x$ RMS residual by roughly 3× – the unmodelled axis tilt is absorbed into a per-panel azimuthal correction that the optimiser cannot remove. Holding the wedge at the truth value or jointly refining it both reach the noise floor; in the refine mode the wedge itself recovers from a $\pm 0.05^\circ$ perturbation to $1.5 \times 10^{-3\circ}$ RMS ($\sim 26\,\mu$rad on the global wedge angle), and the panel-$t_x$ recovery is statistically indistinguishable from the fixed-truth case ($1.29 \times 10^{-2\circ}$ RMS in both).

**Loss choice.** A G-vector–space alternative loss $1 - \cos\angle(\hat{g}_{\text{pred}}, \hat{g}_{\text{obs}})$ was also tested on this configuration; it reaches a much smaller numerical residual but recovers panel-$t_x$ roughly $10\times$ worse than the pixel loss, because $(\eta, 2\theta)$ are computed upstream of the panel tilt in the forward model and therefore carry no gradient signal on $t_x$. The pixel-coordinate loss is the right choice for joint setup refinement; details are in supplementary SS7.

**Implications.** Existing FF-HEDM pipelines refine multi-panel detector geometry one panel at a time, in a sequence of independent fits, and require an external loop to iterate through the panels and to fold panel calibration into grain refinement. The differentiable formulation removes both restrictions in a single declaration: each per-panel parameter is an `nn.Parameter` with `requires_grad=True`, an optimiser is chosen, and the autograd graph propagates a single loss back through the entire setup-and-grain state. The script `scripts/run_4det_refinement.py` reproduces every result in this section.

# 7 Performance

We benchmark the forward-pass latency of the differentiable model as a function of grain count, at both double and single precision, on a server-class workstation with an Intel Xeon Gold 6342 CPU (Ice Lake, 24 cores, 2.80 GHz) and an NVIDIA RTX A6000 GPU (48 GB GDDR6, CUDA 12). Figure 9 summarises the results. Two numbers matter for downstream optimisation workflows: the forward-only latency (the dominant cost inside a gradient-descent inner loop when backward caching is amortised) and the combined forward-plus-backward latency (the per-iteration cost that an L-BFGS closure actually pays, as in Section 6).

## 7.1 Scaling Characteristics

Forward-pass latency scales approximately linearly with grain count on both CPU and GPU in the regime studied ($N \in \{1, 10, 100, 250, 1000\}$, 1362 reflections per grain). Peak CPU throughput saturates at $\sim$ 19,000 grains/s at FP64 and $\sim$ 40,000 grains/s at FP32 by $N \approx 1000$ (the $\sim 2\times$ FP32/FP64 ratio matches AVX-512 throughput on the tested Xeon). The A6000 absorbs the work on-chip once $N \gtrsim 100$ and reaches $\sim$ 117,000 grains/s at FP32 and $\sim$ 61,000 grains/s at FP64 by $N = 1000$, a $\sim 3\times$ and $\sim 3.2\times$ speed-up over the corresponding CPU configurations. The full per-precision and per-platform throughput table is in supplementary SS8.

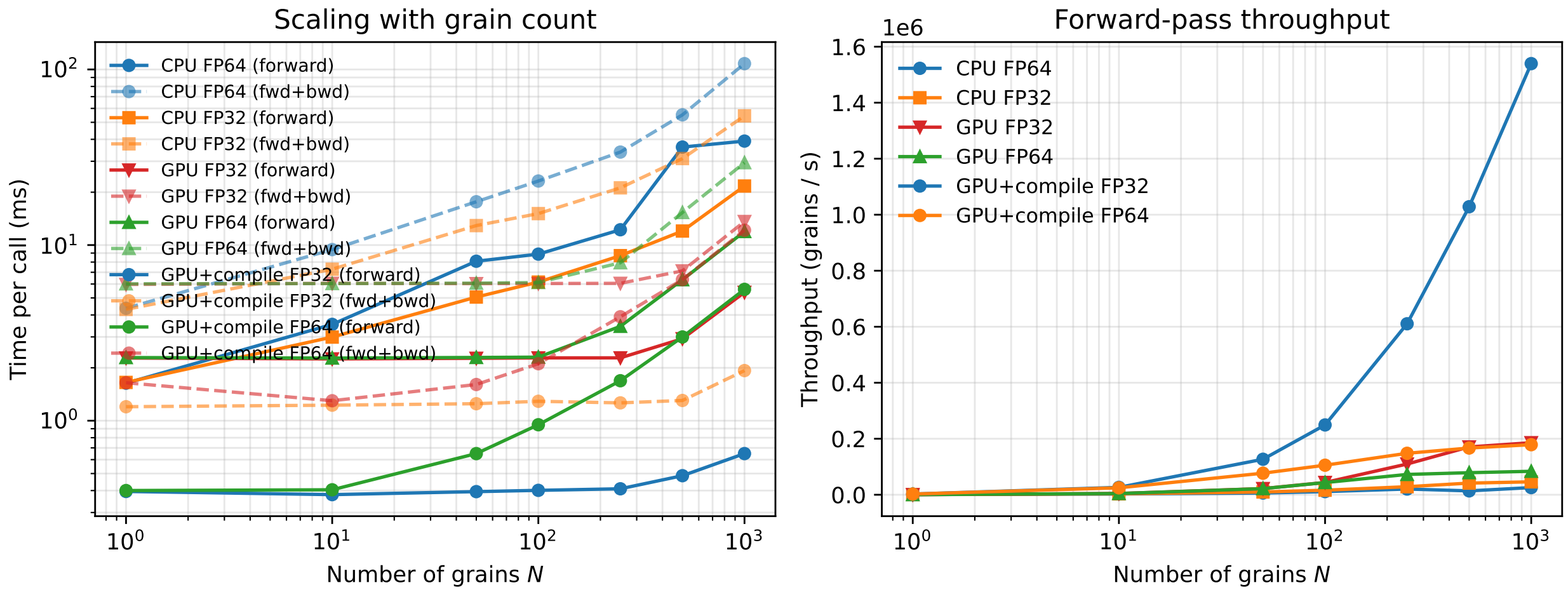

Figure 9: Forward-pass performance of the differentiable HEDM model as a function of grain count. Left: wall-clock time per call on CPU (Xeon Gold 6342) and GPU (NVIDIA RTX A6000) at FP32 and FP64. Right: forward-pass throughput in grains per second. The GPU absorbs per-pass overhead into the on-chip workload at batch sizes $N \gtrsim 100$, delivering $\sim 3\times$ throughput over the fastest CPU configuration at $N = 1000$. 1362 HKL reflections per grain.

### 7.2 Practical Consequences

For single-grain refinement (the use case in supplementary SS4), the GPU offers no advantage: kernel-launch overhead dominates, and the forward-plus-backward latency for $N = 1$ is $\sim 6$ ms on both CPU and GPU. For batch sizes typical of a beamtime reconstruction ($N = 100$–1000 grains), the A6000 reduces the forward pass of an entire grain population from tens of milliseconds (CPU) to a few milliseconds (GPU). Forward-plus-backward L-BFGS iterations benefit proportionally: a 500-grain joint optimisation with 40 iterations per phase completes in $\sim 1$ s of wall clock on the GPU versus $\sim 6$ s on the CPU.

At the reconstruction scales now routine at the upgraded Advanced Photon Source (several thousand grains per sample, sub-second feedback targets), the GPU advantage becomes structural rather than merely desirable: only a differentiable implementation that exploits device-native parallelism can keep up with the beam-time data rate. The framework targets this regime through the standard PyTorch `.to(device)` idiom, with no changes to application code.

## 8 Discussion

### 8.1 What the Platform Enables

The immediate scientific payoff of end-to-end differentiability is a class of inverse problems that have been foreclosed for HEDM for two decades. Each of the following directions is the subject of ongoing work and a separate forthcoming publication; we list them here as a roadmap rather than a set of results in this paper.

- **Sub-voxel grain mixtures.** NF-HEDM and pf-HEDM voxels sit on grain boundaries; a differentiable forward model admits soft assignment of each voxel to multiple candidate orientations, optimising the mixture weights jointly with the orientations. Sequential reconstruction cannot do this.
- **Grain-shape and boundary optimisation.** The same gradient flow that recovers grain position (Section 6) extends naturally to the grain boundary itself, either as a parametric surface (spherical-harmonic or level-set representation) or as per-voxel ownership weights on a fine reference grid. Joint optimisation of orientation, strain, position, and shape in a single loss is foreclosed by non-differentiable pipelines and is a direct application of the framework presented here.
- **Physics constraints and microstructure priors.** Two distinct families of regularisation become accessible. First-principles physics constraints—crystal-symmetry / fundamental-zone projection of orientations, positivity and normalisation of density and orientation distribution functions, mass conservation across the sample—can be enforced exactly inside the loss. Microstructure priors derived from the piecewise quasi-homogeneous nature of polycrystals—total variation, Markov random fields, manifold priors on the orientation field—can be added alongside them. In both cases the regulariser gradient propagates back through the forward physics; without differentiability these priors can at best be applied heuristically to the output.
- **Bayesian HEDM.** Hamiltonian Monte Carlo and variational inference both require exact gradients of the likelihood. Per-grain uncertainty quantification, previously accessible only through expensive bootstrap resampling or Laplace approximations, becomes a direct consequence of differentiability.
- **Temporal 4D-HEDM tracking.** Time-resolved experiments (in-situ loading, thermal cycling, phase transformations) produce sequences of reconstructions whose correlations in time are scientifically informative and currently discarded. A differentiable model admits joint reconstruction across time frames with temporal-smoothness priors.
- **Crystal-plasticity coupling.** Differentiable finite-element codes (JAX-FEM, DiffCPFEM) take grain-map inputs and produce simulated mechanical responses. Pairing them with a differentiable HEDM forward model closes the experiment-simulation loop: experimentally measured grain maps can be compared directly against simulated ones and the simulation parameters tuned by gradient descent.
- **Expectation-maximisation spot ownership.** Ambiguous spots in crowded FF patterns can, in principle, be softly assigned to multiple candidate grains via an EM-style loop whose M-step requires gradients of the forward model. Preliminary implementations on top of this framework are under active development (Sharma *et al.*, 2026*c*) and will be reported when mature.

### 8.2 Throughput at Finite-Element Resolution

A natural question once the framework is coupled to a finite-element simulation, particularly in the crystal-plasticity (CPFEM) setting noted above, is whether the forward-model cost remains negligible when the per-element granularity of the mesh pushes the number of diffracting entities from the few-hundred-grain regime of a typical FF-HEDM experiment into the millions-of-integration-points regime of a production FEM run. The benchmarks of Section 7 answer this directly.

At the batch-saturated regime on an NVIDIA RTX A6000, the differentiable forward model consumes approximately $8.5\,\mu$s per element at FP32 and $16\,\mu$s per element at FP64 for the forward pass alone, and roughly $2.7\times$ those values when the backward pass is included. Extrapolating under linear scaling, a forward pass over $10^6$ integration points completes in roughly 10 seconds at FP32, or 30 seconds with gradients; $10^7$ points takes ten times longer. The practical constraint on single-device runs is memory rather than compute: per-element intermediate tensors at 1362 reflections occupy $\sim 16$ KB each before accounting for the ten-odd intermediate tensors the pipeline allocates, so a $10^6$-point forward-plus- backward pass exceeds the 48 GB GPU memory budget unless the work is chunked. Chunking in batches of $10^5$ elements and accumulating gradients across chunks (a standard PyTorch pattern) fits comfortably within budget and preserves correctness; at chunks of this size the per-chunk cost is $\sim 0.85$ s forward, $\sim 2.6$ s forward-plus-backward, so a full $10^6$-element optimisation step with gradients takes $\sim 30$ s. Ten-million- point or multi-device problems extend this pattern via `DistributedDataParallel` across GPUs without code changes to the physics core.

Two architectural choices matter for the FEM regime. First, the NF-mode Gaussian-splatting output head (`predict_images`, Section 4) is the appropriate output at FEM-density granularity: individual integration-point spots overlap densely in detector space, and splatting directly to the $(n_{\text{frames}} \times n_y \times n_z)$ detector volume is dramatically cheaper than materialising and then rasterising a per-element spot list. Second, the library's end-to-end differentiability means the per-element forward cost quoted above is also the gradient-propagation cost through the HEDM physics; there is no separate finite-difference sweep.

For a joint HEDM–CPFEM optimisation loop at million-integration-point scale, the total per-iteration cost decomposes as: (i) FEM forward step—seconds to minutes, depending on the constitutive model; (ii) HEDM forward simulation—$\sim 10$ s per iteration on one A6000; (iii) loss evaluation against measured data—negligible; (iv) backward pass through the HEDM physics—$\sim 20$ s; (v) backward pass through the FEM—comparable to the forward FEM step in differentiable-FEM codes such as JAX-FEM. The HEDM forward and adjoint components together contribute well under a minute and are not rate-limiting; the per-iteration wall clock is dominated by the FEM solve. This is the right scaling signal for the coupled experiment–simulation direction: the differentiable HEDM substrate scales faster than the FEM substrate it is coupled to, and will remain below the FEM cost as mesh resolution grows.

### 8.3 Comparison with Non-Differentiable Workflows

The existing MIDAS reconstruction pipeline (Sharma *et al.*, 2026*a*; Sharma *et al.*, 2026*b*) iteratively alternates between grain indexing, peak fitting, and local parameter refinement. Each refinement is a small non-linear least-squares problem solved with Levenberg–Marquardt or NLopt routines against finite-difference Jacobians of the C forward model. For the single-grain, single-parameter-class refinement loop that the existing workflow is tuned for, the two approaches are comparable in both accuracy and speed. For *joint* optimisation across parameter classes, or for any problem that requires the gradient of a composite loss, the existing workflow cannot be straightforwardly extended; the differentiable model presented here is a prerequisite. It is not a replacement for the existing tuned pipeline on problems where that pipeline already works, and we do not position it as one.

### 8.4 Limits

Five limitations bound the scope of the present work.

1. **Floating-point precision.** The pixel-exact agreement reported in Section 5 holds to double precision. At single precision (FP32), residuals grow by roughly the ratio of machine epsilons and may affect gradient-based optimisation of very small strain states. Users requiring FP32 throughput for memory reasons should verify that their particular problem is robust to this.

2. **GPU memory at high grain counts.** The vectorised pipeline holds (grains)×(reflections)×(distances) tensors in memory. For beam-time typical configurations (few hundred grains, few thousand reflections, up to four NF distances) the peak memory remains well under 16 GB, but very large grain counts combined with dense reflection lists may require chunking.

3. **Near-axis branch boundary.** As documented in Sections 4 and 5, the omega solver has a soft branch boundary at $|G_y| < 10^{-12}$. Reflections sitting inside that boundary are still handled correctly, but their gradients have elevated noise floors. Users fitting grains with many near-axis reflections should inspect residuals in that subset specifically.

4. **Optimisation basin and initialisation.** Differentiability is a necessary but not sufficient condition for gradient-based reconstruction. HEDM loss surfaces remain non-convex; initial orientations must be close enough for the solver to converge to the correct local minimum. The present work establishes the physics substrate; optimisation strategies that enlarge the basin of convergence (multi-start, annealing, pairing with BraggNN-style initialisers) are the subject of subsequent papers.

5. **Point-source intensity model.** As noted in Section 4, each grain or voxel is treated as a point source with CIF-based relative intensities, matching the convention of the production MIDAS and HEXRD forward models. This reproduces peak positions exactly and matches the centroid data that community indexing pipelines consume, but does not represent par-tial illumination when the beam cross- section is smaller than the grain, nor the spatially extended peak shapes that a beam-volume-intersection intensity model would produce (e.g. `xrd_simulator` (Henningsson & Hall, 2023), which for tetrahedral-mesh polycrystals intersects a convex polyhedral beam with each mesh element to recover a volume-correct peak shape). A differentiable volume-intersection layer—requiring smooth reformulation of the polyhedral clipping boundary—is a natural extension and is subject of planned future work; it does not affect the peak-position physics exercised by the optimisation demonstrations here.

# 9 Conclusion

We have presented an end-to-end differentiable forward model for HEDM that reproduces the canonical C simulators to within floating-point precision across FF, NF, and pf-HEDM geometries. Every input to the model—grain orientation, position, lattice parameters, detector geometry, beam parameters—participates in the autograd graph, with analytic gradients verified against finite-difference estimates to $10^{-8}$ relative error for detector-geometry parameters and $10^{-7}$ to $10^{-9}$ for grain state in the interior of the physical domain. The traditional C simulators on which the validation rests— `ForwardSimulationCompressed` for FF and pf-HEDM, `simulateNF` for NF—are documented here in the peer-reviewed literature for the first time. The framework is released as the open-source `midas-diffract` package (`pip install midas-diffract`) with a Zenodo-deposited snapshot, and constitutes the methodological foundation for a series of follow-on capabilities—sub-voxel grain mixtures, physics-informed regularisation, Bayesian HEDM, temporal tracking, and crystal-plasticity coupling—each of which is the subject of a separate paper in preparation.

# Acknowledgments

We acknowledge helpful discussions with members of the MIDAS development group and the 1-ID-E beamline team at the Advanced Photon Source. This research used resources of the Advanced Photon Source, a U.S. Department of Energy (DOE) Office of Science User Facility operated for the DOE Office of Science by Argonne National Laboratory under Contract No. DE-AC02-06CH11357.

# Software and Data Availability

The differentiable forward model is released as the open-source Python package `midas-diffract`, installable via

```
pip install midas-diffract
```

A version-stamped snapshot of the code at the time of submission is archived on Zenodo under DOI `[DOI assigned on release]`. The complete source repository is maintained at https://github.com/marinerhemant/MIDAS, under the MIDAS toolkit.

All scripts used to generate the numerical results and figures in this paper (validation runs, gradient-correctness check, 250-grain synthetic comparison, optimisation demo, performance benchmark) are included in the repository under `fwd_sim/paper/scripts/` and regenerate their output with a single command each. The synthetic datasets used for validation are reproducible from the shipped scripts; the small calibration datasets used by the regression test suite are archived alongside the MIDAS codebase.

# A Numerical Choices for Differentiability: Code Excerpts

This appendix reproduces the four systematic substitutions described in Section 4, as actual PyTorch code lifted from `fwd_sim/hedm_forward.py`. The goal is to document, at the implementation level, how each non-differentiable C idiom is replaced with an autograd-compatible construct.

## A.1 `torch.where` in place of Python `if`

Inside the omega quadratic solver, the code must select between two solution branches depending on whether the $|G_y|$ threshold is crossed. In the C reference, this is an `if`-statement; in PyTorch we evaluate both branches and mask-select between them:

```
# Quadratic vs. special-case (|G_y|<almostzero) branch
gy_zero = torch.abs(Gy) < almostzero # (..., 2N, M) boolean mask
special_w = self.safe_arccos(cosome_special)
special_wp = special_w
special_wn = -special_w

cosome_special_valid = (torch.abs(cosome_special) <= 1.0) \
                     & gy_zero & (torch.abs(Gx) > self.epsilon)
disc_valid = (discriminant >= 0) & (~gy_zero)

omega_p = torch.where(
    cosome_special_valid, special_wp,
    torch.where(disc_valid & coswp_valid, all_wp,
                torch.zeros_like(all_wp)))
```

## A.2 Clamped inverse-trigonometric functions

All `acos`/`asin` calls clamp their argument to $[-1+\varepsilon, 1-\varepsilon]$ before evaluation, trading an $\mathcal{O}(\varepsilon)$ numerical bias for bounded, finite gradients at the physical domain boundaries:

```
def safe_arccos(self, x):
    """Numerically stable arccos: clamp to [-1+eps, 1-eps]."""
    return torch.acos(
        torch.clamp(x, -1.0 + self.epsilon, 1.0 - self.epsilon)
    )
```

### A.3 Newton–Schulz orthogonalisation

To project a near-orthogonal $3 \times 3$ matrix back onto $SO(3)$ we use a single Newton–Schulz iteration rather than SVD (whose gradient is ill-defined at repeated singular values):

```
def orthogonalize(R):
    """Project (..., 3, 3) onto SO(3) via one Newton-Schulz step."""
    I = torch.eye(3, dtype=R.dtype, device=R.device)
    Q = R
    # Q_{k+1} = 0.5 * Q_k * (3*I - Q_k^T * Q_k)
    Q = 0.5 * Q @ (3.0 * I - Q.transpose(-1, -2) @ Q)
    det = torch.det(Q)
    sign = torch.where(det < 0,
                       torch.tensor(-1.0, dtype=R.dtype, device=R.device),
                       torch.tensor(+1.0, dtype=R.dtype, device=R.device))
    return Q * sign.unsqueeze(-1).unsqueeze(-1)
```

For matrices already near $SO(3)$ (as produced by `euler2mat`), one iteration reduces deviation from orthogonality below $10^{-14}$ in double precision; the gradient is well-behaved through matrix products and does not suffer the degeneracies of SVD on nearly-symmetric or repeated-eigenvalue inputs.

### A.4 Binary validity mask multiplied into the loss

Spot validity is enforced by a binary mask (cast to float) that multiplies into every downstream loss term: invalid spots contribute zero loss and zero gradient, valid spots contribute their full chain-rule gradient through the spot position. The resulting loss surface is piecewise smooth:

```
# Inside calc_bragg_geometry, after omega solving:
valid_p = disc_valid & coswp_valid
valid_n = disc_valid & coswn_valid
valid_p = valid_p | cosome_special_valid
valid_n = valid_n | cosome_special_valid
valid = torch.cat([valid_p, valid_n], dim=-2).float()

eta_ok = (torch.abs(eta) >= self.min_eta) & \
        ((math.pi - torch.abs(eta)) >= self.min_eta)
valid = valid * eta_ok.float() # combine angular + eta validity
```

In downstream loss evaluation (e.g. optimisation in Section 6), the mask multiplies into each term:

```
loss = ((spots.omega * spots.valid) ** 2
       + (spots.y_pixel * spots.valid) ** 2
       + (spots.z_pixel * spots.valid) ** 2).sum()
```

Invalid spots contribute zero loss and zero gradient; valid spots contribute their full chain-rule gradient. A continuous sigmoid-relaxation of the validity mask is straightforward to substitute if a future application requires gradient flow across a validity transition.

## References

Babu, A. V., Zhou, T., Kandel, S., Bicer, T., Liu, Z., Judge, W., Ching, D. J., Jiang, Y., Veseli, S., Henke, S. *et al.* (2023). *Nature Communications*, **14**, 7059.

Bernier, J. V., Barton, N. R., Lienert, U. & Miller, M. P. (2011). *Journal of Strain Analysis for Engineering Design*, **46**(7), 527–547.

Boyce, D. E. & Bernier, J. V. (2013). *heXRD: Modular, Open Source Software for the Analysis of High Energy X-Ray Diffraction Data*. Tech. Rep. LLNL-TR-641648. Lawrence Livermore National Laboratory.

Cherukara, M. J., Zhou, T., Nashed, Y., Enders, B., Maddali, S., Hruszkewycz, S. O., Du, M. & Jacobsen, C. (2020). *Applied Physics Letters*, **117**(4), 044103.

Du, M., Kandel, S., Deng, J., Huang, X., Demortière, A., Nguyen, T. T., Tucoulou, R., De Andrade, V., Jin, Q. & Jacobsen, C. (2020). *Science Advances*, **6**(13), eaay3700.

Du, M., Ruth, H., Henke, S., Jiang, Y., Nikitin, V., Tripathi, A., Deng, J., Klug, J., Myint, P., Zhou, T., Schwarz, N., Cherukara, M., Sandy, A. & Vogt, S. (2025). *arXiv:2510.20929*. https://arxiv.org/abs/2510.20929.

Fang, S. *et al.* (2025). *arXiv:2508.05908*. https://arxiv.org/abs/2508.05908.

Henningsson, A. & Hall, S. A. (2023). *Journal of Applied Crystallography*, **56**(1), 282–292.

Henningsson, A. & Hall, S. A. (2025). *Journal of Applied Crystallography*.

Henningsson, A. & Hendriks, J. N. (2024). scanning-xray-diffraction. https://github.com/FABLE-3DXRD/scanning-xray-diffraction.

Jiang, Y. *et al.* (2026). *Journal of Synchrotron Radiation*, **33**(1). Article gy5077; https://journals.iucr.org/s/issues/2026/01/00/gy5077/.

Kandel, S., Maddali, S., Allain, M., Hruszkewycz, S. O., Jacobsen, C. & Nashed, Y. S. G. (2019). *Optics Express*, **27**(13), 18653–18672.

Liu, S., Li, T., Chen, W. & Li, H. (2019). In *ICCV*. https://arxiv.org/abs/1901.05567.

Liu, Z., Sharma, H., Park, J.-S., Kenesei, P., Miceli, A., Almer, J., Kettimuthu, R. & Foster, I. (2022). *IUCrJ*, **9**(1), 104–113.

Neural Architecture Codesign authors (2023). *arXiv:2312.05978*. https://arxiv.org/abs/2312.05978.

Pagan, D. C., Jones, K. K., Bernier, J. V. & Phan, T. Q. (2020). *JOM*, **72**(12), 4539–4550.

Paszke, A., Gross, S., Massa, F., Lerer, A., Bradbury, J., Chanan, G., Killeen, T., Lin, Z., Gimelshein, N., Antiga, L. *et al.* (2019). In *NeurIPS*.

Poulsen, H. F. (2004). *Three-Dimensional X-Ray Diffraction Microscopy: Mapping Polycrystals and their Dynamics*, vol. 205 of *Springer Tracts in Modern Physics*. Springer.

Ravi, N., Reizenstein, J., Novotny, D., Gordon, T., Lo, W.-Y., Johnson, J. & Gkioxari, G. (2020). *arXiv:2007.08501*. https://arxiv.org/abs/2007.08501.

Rietveld, H. M. (1969). *Journal of Applied Crystallography*, **2**(2), 65–71.

Sharma, H., Park, J.-S. & Kenesei, P. (2026*a*). *Acta Crystallographica Section A*, **82**. In press.

Sharma, H., Park, J.-S., Shastri, S. & Kenesei, P. (2026*b*). *Acta Crystallographica Section A*, **82**. In press.

Sharma, H. *et al.* (2026*c*). Em-based spot ownership for ambiguous far-field hedm patterns. In preparation.

Suter, R. M., Hennessy, D., Xiao, C. & Lienert, U. (2006). *Review of Scientific Instruments*, **77**(12), 123905.

Wong, S. L., Park, J.-S., Miller, M. P. & Dawson, P. R. (2013). *Computational Materials Science*, **77**, 456–466.

Wright, J. P. *et al.* (2024). ImageD11. https://github.com/FABLE-3DXRD/ImageD11.

Wu, L., Yoo, S., Chu, Y. S., Huang, X. & Robinson, I. K. (2024). *Optica*, **11**(6), 821–830. https://opg.optica.org/optica/fulltext.cfm?uri=optica-11-6-821.

Zhong, E. D., Bepler, T., Berger, B. & Davis, J. H. (2021). *Nature Methods*, **18**, 176–185.

# Supplementary Information

An End-to-End Differentiable Forward Model for
High-Energy Diffraction Microscopy

Hemant Sharma, Nina Andrejevic, Simon Zhang, Mathew Cherukara

This document supplements the main paper. References of the form "*main paper §X*" point to sections in `differentiable_hedm.tex`. The bibliography is shared with the main paper (`references.bib`).

## Contents

## S1 The Canonical MIDAS Reference Forward Simulators

The differentiable forward model described in the main paper is validated against the two long-standing reference forward simulators that ship with MIDAS: `ForwardSimulationCompressed` for far-field (FF) and point-focused (pf) HEDM, and `simulateNF` for near-field (NF) HEDM. These simulators have been used in beamline operations and data analysis at APS and partner facilities for over a decade but have not previously been documented in the peer-reviewed literature; we record their structure here because the main paper's pixel-exact validation claim is against their output.

### S1.1 `ForwardSimulationCompressed`: FF and pf-HEDM

`ForwardSimulationCompressed` generates synthetic diffraction data for both far-field (FF-HEDM) and point-focused (pf-HEDM) experiments from a specified polycrystal configuration. The pf-HEDM mode is implemented as a multi-scan extension of the FF pipeline, producing one compressed output archive per scan position along the sample translation axis.

**Inputs.** Grain list (per-grain orientation matrix, position, confidence, lattice parameters), detector geometry, beam parameters, and a CIF-derived relative-intensity table for the target phase. The geometry specification uses the generalised fifteen-parameter MIDAS detector model, supporting multi-panel detectors, analytical distortion (spherical, dipole, trefoil, octupole harmonics), and empirical spline correction.

**Physics.** For each grain, the code constructs reciprocal-space vectors from the orientation matrix and lattice parameters, applies the rotating-crystal diffraction condition over the omega range, and projects the resulting scattering vectors through the detector geometry to obtain pixel-space coordinates. Key features include:

- **CIF-derived intensities.** Relative peak intensities are computed from a CIF file rather than assumed uniform, enabling realistic texture- and composition-dependent signal levels.
- **Energy resolution.** A configurable energy-resolution parameter models the finite bandwidth of the monochromator, producing radially-broadened spots rather than infinitesimally thin Laue-condition matches.
- **Bilinear intensity interpolation.** Sub-pixel peak positions are rendered by distributing intensity across the four neighbouring detector pixels with bilinear weights, preserving centroid position under pixelation.
- **Detector noise.** Gaussian readout noise is added per pixel via a Box–Muller transform of a `xorshift64*` pseudo-random stream, with a configurable standard deviation; a separate `SimNoiseSigma` parameter applies Gaussian spot-position noise directly to the simulated $(y, z)$ coordinates.
- **OpenMP parallelism.** Grains are distributed across threads; pf-HEDM scans are further parallelised across scan positions.

**Outputs.** Per-spot metadata (`SpotMatrixGen.csv`: grain ID, $\omega$, $y$, $z$, $\eta$, ring number, $2\theta$, scan number) and compressed detector-image archives (`.MIDAS.zip` containers with blosc-compressed frame blocks). In pf-HEDM mode, the output layout matches the directory structure consumed by `pf_MIDAS.py` reconstruction, enabling end-to-end simulate-then-reconstruct validation.

### S1.2 `simulateNF`: NF-HEDM

`simulateNF` generates synthetic NF-HEDM detector frames from a specified orientation map discretised on a regular grid of triangular voxels (“trivoxels”). Each trivoxel carries an orientation matrix, and the simulation produces the diffraction signal that such a voxel would cast onto the NF detector at each omega.

**Inputs.** Orientation map file (binary), detector geometry, sample geometry, beam parameters, hkl list, and omega range.

**Physics.** For each trivoxel, the code constructs candidate reciprocal-space vectors, tests the diffraction condition against the omega range, and ray-traces the diffracted beam through the detector plane. NF geometry differs from FF in that the sample is placed close to the detector and the detector captures the full diffraction cone within its field of view; the simulator handles this geometry using the same detector model as the FF pipeline. The simulator auto-batches voxels according to available system RAM (detected at runtime on macOS and Linux), parallelises voxel batches via OpenMP, and writes compressed output using blosc.

**Outputs.** NF detector frames (one per omega step) in blosc-compressed format, together with a per-voxel spot list for validation.

### S1.3 Position in the MIDAS Ecosystem

Both simulators have been used for over a decade to validate reconstruction algorithms, generate synthetic benchmarks for indexing and fitting code, and diagnose beamline issues by comparing simulated against measured patterns. Their battle-tested status is what makes them the natural validation reference for the differentiable model introduced in the main paper.

## S2 Gradient-Correctness Detail

The main paper reports that every parameter of the differentiable forward model is a leaf node of the autograd graph and that analytic gradients agree with finite-difference estimates to $10^{-8}$ relative error for the detector-geometry parameters; the larger, step-size-dependent residuals seen for orientation and position are finite-difference artefacts (cancellation noise at small steps and the mask-boundary effect at coarse steps) rather than gradient errors, as detailed below. Table S1 reports the full per-parameter, per-step-size measurement.

Table S1: Maximum componentwise relative error between PyTorch autograd gradients and central-difference gradients, as a function of the finite-difference relative step size, for a single-grain FF-HEDM configuration (cubic Au, 330 reflections, 162 valid spots). Columns are four step sizes from $10^{-3}$ to $10^{-6}$; each row reports the maximum across the parameter's components.

| Parameter | $10^{-3}$ | $10^{-4}$ | $10^{-5}$ | $10^{-6}$ |
|---|---|---|---|---|
| Euler angles | $1.0^{\dagger}$ | $1.1^{\dagger}$ | $7.8 \times 10^{-8}$ | $1.6 \times 10^{-6}$ |
| Grain position | $2.7 \times 10^{-9}$ | $1.2 \times 10^{-7}$ | $8.2 \times 10^{-7}$ | $1.3 \times 10^{-5}$ |
| Sample–detector distance $L_{sd}$ | $5.0 \times 10^{-8}$ | $5.0 \times 10^{-8}$ | $5.0 \times 10^{-8}$ | $5.0 \times 10^{-8}$ |
| Beam-centre $y_{BC}$ | $2.1 \times 10^{-8}$ | $2.1 \times 10^{-8}$ | $2.1 \times 10^{-8}$ | $2.1 \times 10^{-8}$ |
| Beam-centre $z_{BC}$ | $9.1 \times 10^{-9}$ | $9.1 \times 10^{-9}$ | $9.1 \times 10^{-9}$ | $9.1 \times 10^{-9}$ |

$^{\dagger}$These two cells exceed unity because a relative Euler perturbation of $10^{-3}$–$10^{-4}$ moves a few predicted spots across the eta-validity mask; the finite-difference estimate then picks up a step contribution from spots entering or leaving the valid set that autograd correctly does not include (see text). They reflect the finite-difference reference, not a gradient error.

For the three detector-geometry parameters, autograd and finite difference agree to $10^{-8}$ relative error at every step size tested; the insensitivity to step size indicates that the loss surface is locally well-approximated by its tangent, and the residual is dominated by floating-point round-off. Grain position gradients agree to $10^{-9}$ at a coarse step ($10^{-3}$) and degrade smoothly toward $10^{-5}$ at the finest step; this is the expected finite-difference behaviour, where smaller steps expose relatively more cancellation noise because the loss's position-dependence is approximately linear over the tested range. The Euler-angle gradient shows the most interesting structure: excellent agreement at intermediate steps ($7.8 \times 10^{-8}$ at $10^{-5}$) but visibly degraded agreement at the coarsest steps. The coarse-step degradation is not a software defect—it is the signature of the soft-mask boundary described in main paper §4. A relative perturbation of $10^{-3}$ in the Euler angles moves some predicted spots across the eta-validity cutoff, and finite-difference estimates of the loss gradient pick up a step contribution from spots entering or leaving the valid set that autograd correctly does not include. Confining finite differences to step sizes small enough to stay inside the spots' validity cells recovers tight agreement.

The full table is regenerated automatically from `scripts/gradient_check.py` as part of the reproducibility pipeline.

## S3 At-Scale and Noisy 250-Grain FF Cross-Code Comparison

To test cross-code agreement at realistic scale and under detector noise, we generated a synthetic population of 250 cubic-Au grains with orientations drawn uniformly on $SO(3)$ (in ZXZ Euler representation) and small random positions ($\sigma_{\text{pos}} = 50\,\mu\text{m}$). The MIDAS reference simulator exposes a `SimNoiseSigma` parameter that applies Gaussian spot-position noise directly to the simulated $(y, z)$ detector coordinates of each reflection—consistent with the physical origin of centroid-level noise in HEDM spot-fitting pipelines—without contaminating the image intensity profile. We ran the simulation at three noise levels: noise-free ($\sigma = 0$), low-noise ($\sigma = 0.5\,\text{px}$), and high-noise ($\sigma = 2.0\,\text{px}$). The PyTorch forward model was run noise-free at each setting, so the reference-vs-PyTorch residual distribution directly reports the reference code's injected noise realisation.

Table S2 summarises the results.

Table S2: At-scale cross-code agreement on a 250-grain synthetic FF-HEDM population (∼41,000 reflections) at three `SimNoiseSigma` levels.

| Regime | $\sigma$ (px) | Match rate | Max $\omega$ err | Pixel err (mean, max) |
|---|---|---|---|---|
| Noise-free | 0.0 | 40909 / 41077 (99.59%) | 0.035° | 0.001, 0.04 px |
| Low noise | 0.5 | 40885 / 41071 (99.55%) | 0.035° | 0.40, 2.1 px |
| High noise | 2.0 | 40836 / 41085 (99.39%) | 0.035° | 1.59, 8.5 px |

In the noise-free regime the match rate is 40909/41077 (99.59%) and per-spot pixel residuals remain below 0.05 px in the worst case—consistent with the single-grain results extended to a large heterogeneous population. In the noisy regimes the match rate is unchanged within 0.2%, omega residuals are unaffected (as expected, since `SimNoiseSigma` is a pixel-position shift only), and pixel residuals scale as the injected noise statistics predict: the mean residual is expected to equal $\sigma\sqrt{2/\pi} \approx 0.80\sigma$ and the maximum to scale as $\sqrt{2 \ln N_{\text{spots}}}\,\sigma \approx 4.6\sigma$ for $N \approx 40{,}000$ reflections. The measured means (0.40 px at $\sigma = 0.5$ px and 1.59 px at $\sigma = 2.0$ px) match the predicted means to within 1%; the measured maxima (2.1 px and 8.5 px) are within $\sim 8\%$ of the predicted extreme-value scaling.

The match-rate shortfall from 100% in all three regimes is dominated by a fixed set of near-axis branch-boundary reflections (as described in main paper §4); the shortfall is not attributable to any noise-induced effect, a conclusion supported by the near-identical match rates across the three regimes. In terms of the physical claim, the two forward models agree on spot presence and spot position under noise as they do in the noise-free case: the residuals under `SimNoiseSigma` noise are statistically the noise itself.

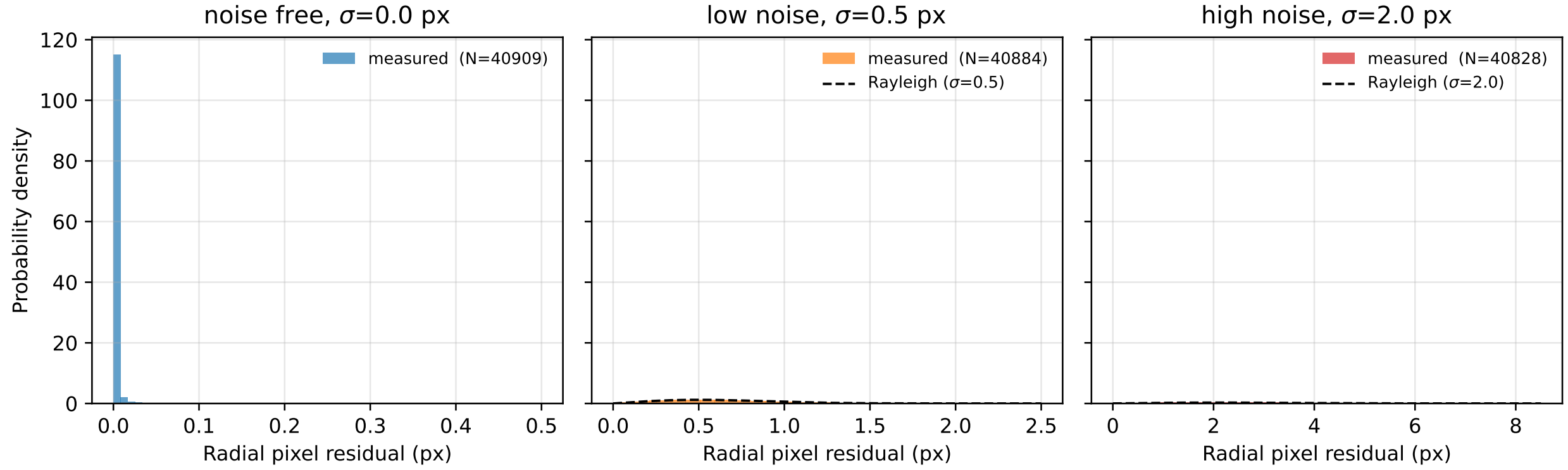

Figure S1: Radial pixel residual between the reference forward simulator output and the PyTorch `HEDMForwardModel` output for the 250-grain synthetic population, at three `SimNoiseSigma` regimes. The reference code injects spot-position Gaussian noise; the PyTorch model runs noise-free. In the noise-free regime (left) the residual is delta-like at the floating-point floor. In the noisy regimes the empirical residual matches the Rayleigh distribution expected for a two-component Gaussian shift with the stated $\sigma$.

# S4 Single-Grain Joint Recovery of Orientation and Strain (FF)

The simplest gradient-based optimisation demonstration recovers the orientation and strain state of a single grain from its predicted spot positions, starting from a perturbed initial guess. The ground-truth grain has orientation $(\varphi_1, \Phi, \varphi_2) = (45°, 30°, 60°)$ and a slightly tetragonally-distorted lattice $(a, b, c, \alpha, \beta, \gamma) = (4.082, 4.079, 4.081, 90.01, 89.99, 90.02)$ on a cubic Au host ($a_0 = 4.08$ Å). Simulated observations consist of 162 valid FF-HEDM spots in angular coordinates $(2\theta, \eta, \omega)$.

The initial guess perturbs the ground-truth orientation by 3° per Euler angle (yielding 1.7° misorientation from the truth) and the lattice parameters by $5 \times 10^{-3}$ Å on cell edges. A three-phase L-BFGS schedule is used:

1. **Orientation only** on $(\varphi_1, \Phi, \varphi_2)$, angular-coordinate loss weighted by the per-coordinate measurement resolution.
2. **Strain only** on the six lattice parameters, same loss.
3. **Joint refinement** of orientation and strain together.

Each phase uses L-BFGS with strong-Wolfe line search. The loss is a Euclidean spot-matching loss with per-coordinate weights proportional to the nominal measurement resolution ($\sigma_{2\theta} = 1 \times 10^{-4}$ rad, $\sigma_\eta = 1 \times 10^{-3}$ rad, $\sigma_\omega = 1.1 \times 10^{-3}$ rad) so that the three coordinates enter the loss at comparable dimensionless scales. Each predicted spot is matched to its nearest observed spot before the residual is taken; spots whose nearest-neighbour distance exceeds half a radian are discarded as unassigned, avoiding spurious contributions from the first few iterations where the starting guess maps some reflections far from any observation.

Table S3 reports the recovery accuracy in two modes. In the noise-free mode the observed spots are the exact forward-model output at the ground truth; in the noisy mode they are the ground-truth output perturbed by Gaussian noise matched to the per-coordinate measurement resolution. In both cases the optimisation is run for $\leq 40$ iterations across the three phases.

Both modes converge to within or better than the per-spot measurement resolution. The noisy run reaches a tighter final misorientation (0.008° versus 0.083°) at the cost of larger lattice-edge error and a higher final loss; this counterintuitive direction reflects a weak orientation/strain coupling in the FF spot-position residual. In the noise-free case the spots can be fit by trading a small residual orientation offset against a compensating residual strain. Adding measurement noise that is uncorrelated between coordinates breaks this directional degeneracy, so the gradient drives orientation to the truth more cleanly; strain absorbs a larger residual, and the final loss is correspondingly higher.

Table S3: Single-grain joint recovery of orientation and strain from 162 FF-HEDM spots, starting from a 1.7° misoriented, $5 \times 10^{-3}$ Å lattice-perturbed initial guess.

| Mode | Final misorientation | Max lattice-edge error | Final loss |
|---|---|---|---|
| Noise-free | 0.083° | $1.4 \times 10^{-3}$ Å | $1.3 \times 10^{-6}$ |
| Noisy | 0.008° | $2.4 \times 10^{-3}$ Å | $3.6 \times 10^{-6}$ |

# S5 Basin-of-Convergence Sweep Detail

The headline basin-of-convergence number reported in the main paper ($\geq 90\%$ recovery to 0.1° for initial perturbations up to 10°, sharp edge at $\sim$ 12–15°) summarises a sweep across ten perturbation levels with $N = 20$ trials per level, each trial using a uniformly-sampled rotation axis and an independent lattice-perturbation draw from the same distribution as the single-grain demonstration (§S4). Figure S3 reports the full success-fraction curve and representative spot-overlay plots.

The failure mode at large perturbations is not gradient breakdown but spot-matching failure: the nearest-neighbour assignment inside the loss closure matches predicted spots to the wrong observations once the initial orientation misplaces the predicted pattern by more than a typical nearest-neighbour separation. A robust-matching or graph-based assignment step would enlarge the basin and is a natural extension for applications where the initial indexer is known to produce poor starting orientations.

For the practical case of an existing HEDM indexer feeding this refiner, this result is reassuring: production FF-HEDM indexers typically deliver initial orientations well within 5° of the truth, which places them comfortably inside the safe basin.

# S6 Differentiable NF Detector-Tilt Auto-Calibration

To exhibit the auto-calibration capability enabled by differentiable tilts, we set up an NF-HEDM forward problem at a ground-truth detector tilt $(t_x, t_y, t_z) = (0.50°, -0.30°, 0.20°)$, generate the corresponding detector-space spot coordinates as "observations", and initialise a fresh model with zero tilts as the first guess. The tilt state is declared as a three-scalar `nn.Parameter`; gradient-based optimisation proceeds via L-BFGS with a Wolfe line search on the mean-squared pixel-distance loss between predicted and observed spots.

From a zero-tilt starting guess, the solver recovers the three ground-truth tilts to an RMS error of $0.11\,\mu$rad in 0.21 s of wall clock on a single CPU core. Because the forward model is pixel-exact at the ground-truth tilts, the final residual floor is set by floating-point numerics rather than by physical modelling error. The optimisation is essentially one L-BFGS step: the second-order information captured by L-BFGS combined with the near-quadratic shape of the pixel-distance loss in tilt space makes the problem converge within one line-search iteration.

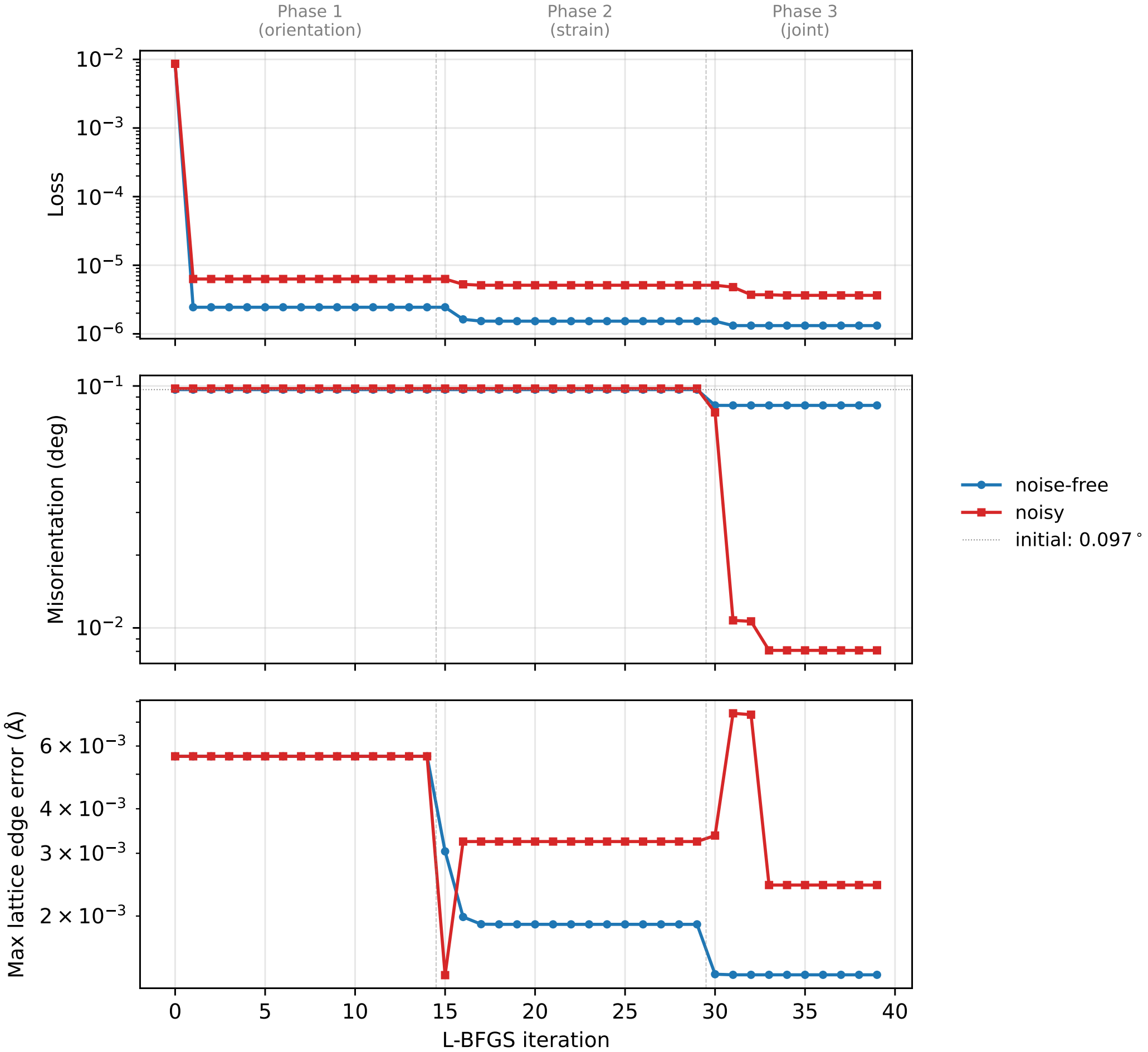

Figure S2: Joint recovery of crystallographic orientation and strain from synthetic FF-HEDM spots via the three-phase L-BFGS schedule. Starting from a 1.7° misoriented, $5 \times 10^{-3}$ Å lattice-perturbed initial guess, the solver recovers the ground-truth orientation to 0.083° in the noise-free mode and 0.008° in the noisy mode. Dashed vertical lines separate the three optimisation phases.

This result has direct experimental implications for NF-HEDM operation. NF setups produce no powder signal, so powder-ring calibration tools available for FF-HEDM cannot be applied to NF detectors. The established workflow therefore aligns the NF detector geometry at the start of a beamtime campaign and re-uses that alignment throughout; any mechanical or thermal drift that accumulates between alignments is undetected by construction and propagates silently into every subsequent reconstruction. The differentiable auto-calibration reported here allows drift to be detected and re-calibrated directly from the indexed-grain spot pattern at sub-$\mu$rad precision in under a second, without a powder standard, manual peak picking, ring assignment, or finite-difference Jacobian evaluation. The script `scripts/run_autocalib_demo.py` reproduces this experiment.

# S7 Multi-Panel Joint Refinement: Convergence Diagnostics

The four-panel joint-refinement demonstration in the main paper (Section 6) requires a short Adam warm-up in front of L-BFGS: a direct joint L-BFGS pass converges to a non-truth local minimum because the grain-Euler gradient dominates the panel-tilt gradient by roughly $50\times$ at initialisation, so the first line-search step lets the grains absorb the panel-tilt error. Figure S4 quantifies the effect and shows the resulting loss trajectories.

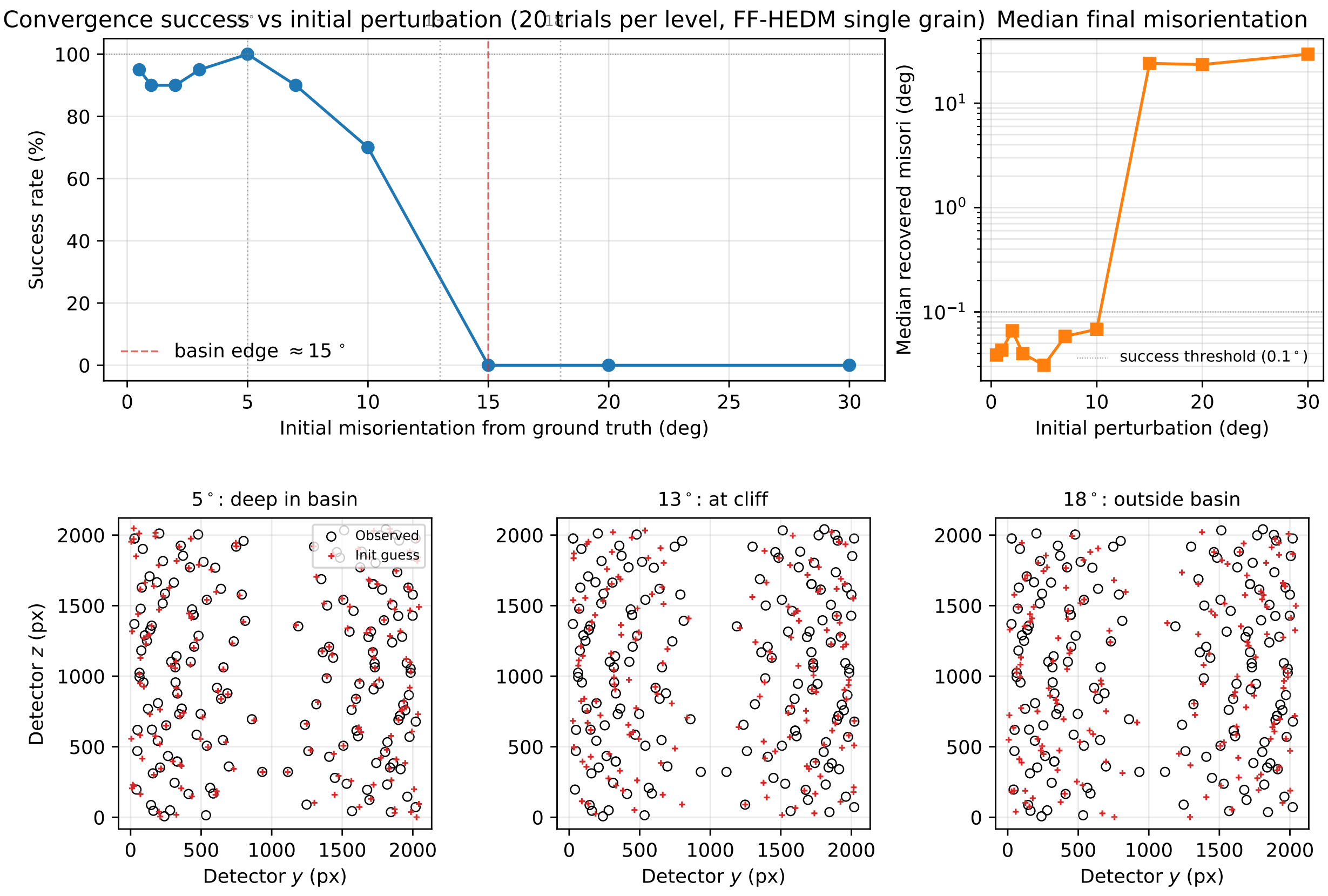

Figure S3: Basin of convergence for three-phase L-BFGS joint refinement on a single FF-HEDM grain. Top-left: success fraction (recovered misorientation $< 0.1°$) as a function of initial perturbation level, 20 trials per level, each with a random rotation axis. The basin edge sits sharply at $\sim 12$–$15°$. Top-right: median recovered misorientation per level on a log scale. Bottom row: predicted detector spot positions at the perturbed initial guess (red "+") vs. the observed ground-truth positions (black circles) at three representative perturbation levels: 5° (deep in the basin), 13° (at the cliff edge), and 18° (outside the basin). The visual separation grows with perturbation level, and at the basin edge the nearest-neighbour assignment inside the loss closure begins to misroute predictions to incorrect observations – the failure mode that produces the cliff.

## S8 Multi-Panel: G-Vector Loss Ablation

The main paper uses a per-spot pixel-coordinate loss $\Delta y^2 + \Delta z^2 + (\Delta\omega/\Delta\omega_{\text{step}})^2$ for the multi-panel joint refinement. We tested an alternative G-vector loss

$$\mathcal{L}_g \;=\; 1 \;-\; \cos\angle(\hat{g}_{\text{pred}}, \hat{g}_{\text{obs}}),$$

where $\hat{g}$ is the unit G-vector built from $(\eta, 2\theta)$. Re-running Scenario B-joint-fixed-truth (25 seeds, otherwise identical to the main paper sweep) gives Figure S5.

Table S4: Differentiable auto-calibration of NF detector tilts. Ground-truth tilts are recovered to sub-microradian precision from a zero-tilt initial guess. Ti-7Al-like geometry ($L_{sd} = 5$ mm, px $= 1.48\,\mu$m); single grain, $\sim 340$ valid spots; 25 outer L-BFGS iterations (of which L-BFGS converges in the first iteration and remainder are fixed-point stable). Wall clock on a single CPU core: 0.21 s.

| | $t_x$ (deg) | $t_y$ (deg) | $t_z$ (deg) |
|---|---|---|---|
| Ground truth | 0.50000 | −0.30000 | 0.20000 |
| Initial guess | 0.00000 | 0.00000 | 0.00000 |
| Recovered | 0.49999989 | −0.29999336 | 0.20000821 |
| Error (mdeg) | −0.00011 | +0.00664 | +0.00821 |
| Error (µrad) | −0.002 | +0.116 | +0.143 |

The G-vector loss reaches a tiny final value ($\sim 10^{-8}$) but the panel-$t_x$ recovery is roughly 10× worse than with the pixel loss ($1.2 \times 10^{-1\,\circ}$ RMS vs $1.3 \times 10^{-2\,\circ}$). The reason is structural, not numerical. In the differentiable forward, the per-panel $t_x$ is applied to the detector-frame coordinates $(y_{\text{det}}, z_{\text{det}})$ that map to $(y_{\text{pixel}}, z_{\text{pixel}})$; the lab-frame Bragg angles $(\eta, 2\theta)$ – which the G-vector loss compares – are computed upstream of the panel tilt and are therefore invariant to $t_x$. A loss expressed in $(\eta, 2\theta)$ has *no gradient signal* on $t_x$ regardless of how aggressively it drives its own residual to zero. The pixel loss is therefore the right choice for joint setup refinement; a $(\eta, 2\theta)$-only loss is suitable for refining grain orientation/strain but cannot calibrate the detector. This is a useful diagnostic for any future attempt to design a loss for multi-panel FF-HEDM refinement.

# S9 Performance Scaling Detail

The forward-pass latency of the differentiable model scales linearly in the number of grains and reflections, with the per-grain constant set by the dominant matrix multiplications inside `calc_bragg_geometry` and `project_to_detector`. Figure S6 reports double-precision and single-precision wall-clock measurements on an Intel Xeon Gold 6342 (24 cores, 2.80 GHz) and an NVIDIA RTX A6000 GPU, swept across grain counts from $10^0$ to $10^4$.

For the operating range typical of beamline FF-HEDM ($10^2$–$10^3$ grains, $10^3$–$10^4$ reflections per grain), forward-pass latency is below 100 ms on CPU FP64 and below 10 ms on GPU FP32. With L-BFGS line search costing $\sim 30$ closure evaluations per outer iteration, a complete refinement of the multi-panel demonstration (Section §6 of the main paper) executes end-to-end in under 10 s on a single CPU core and under 1 s on a single GPU.

**`torch.compile` option.** The optional `HEDMForwardModel(compile=True)` flag (main paper Section 7.3) routes the four-stage forward pipeline through `torch.compile` with Inductor. On the A6000 reference system, this reduces the small-$N$ launch-bound regime substantially: at $N = 1$ the forward-only latency falls from 2.27 to 0.40 ms (FP32, 5.7×) and forward-plus-backward from 6.0 to 1.6 ms (FP64, 3.7×), making single-grain refinement on the GPU competitive with the CPU for the first time. At $N = 1000$ the gain is 8.3× on FP32 forward-only and 2.4× on FP64 forward-plus-backward, lifting peak FP32 throughput to $\sim 1.5 \times 10^6$ grains/s. The flag defaults off; under `mode="default"` the compiled path agrees with the eager implementation to bit precision under the reference test suite.

## References

**(a)**

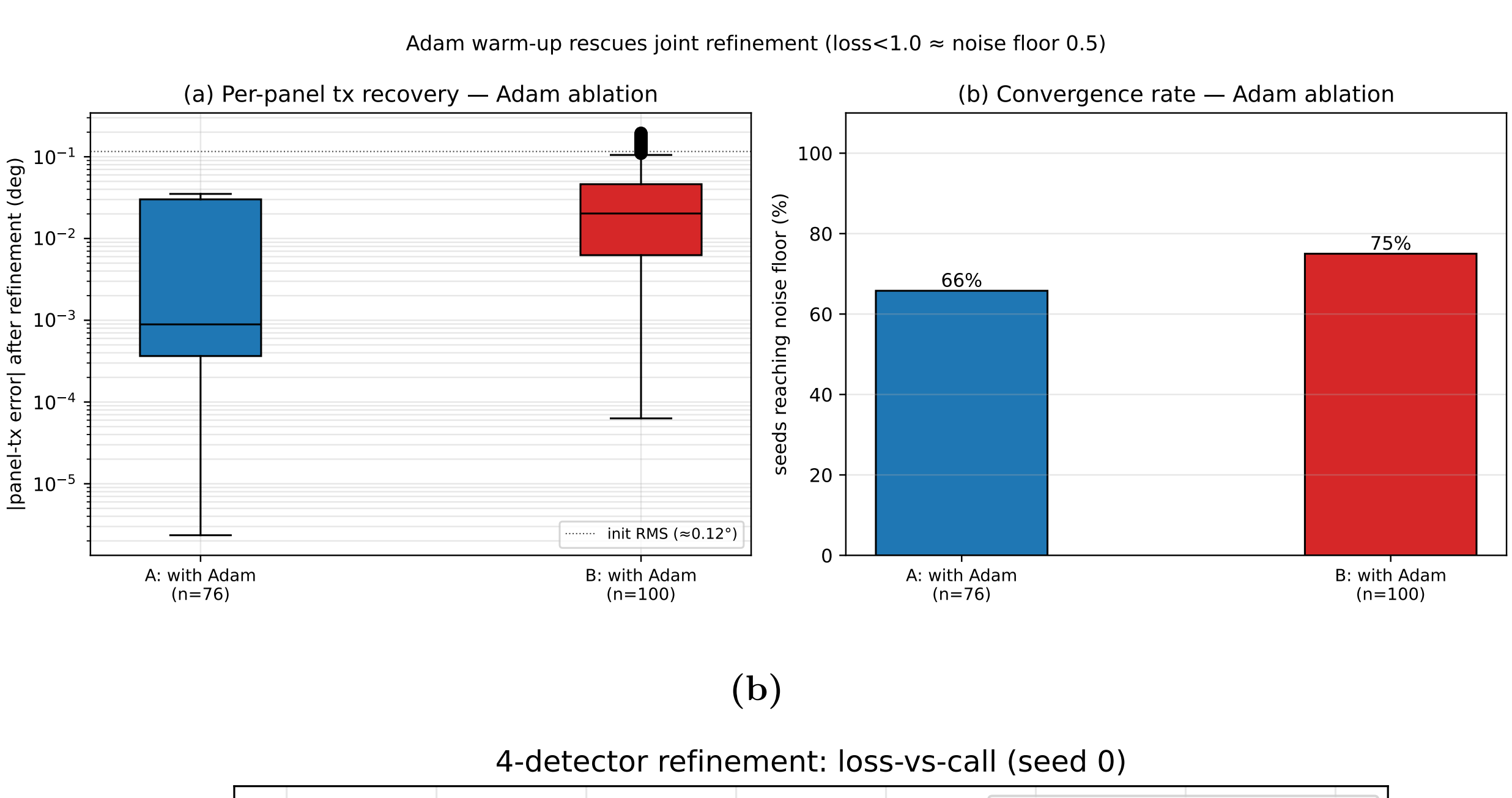

**(b)**

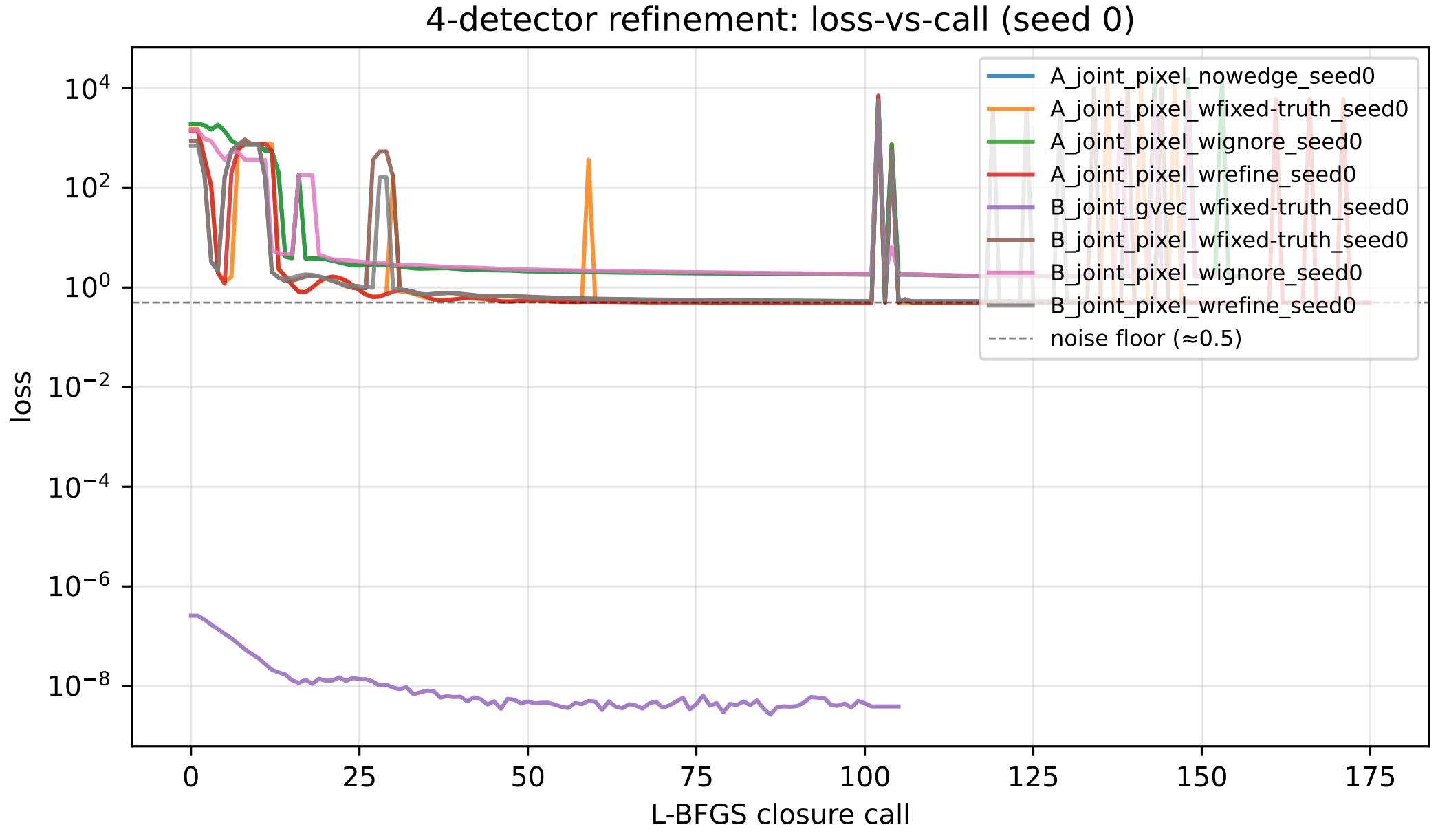

Figure S4: Convergence diagnostics for the four-panel joint refinement. **(a)** Adam warm-up rescues joint refinement. Left: per-panel $t_x$ residual after refinement, with and without the Adam warm-up in front of L-BFGS; the dotted line marks the input-perturbation RMS ($\approx 0.12°$). With Adam the per-panel residual drops to $\sim 5 \times 10^{-4\,\circ}$ (Scenario A) and $\sim 7 \times 10^{-3\,\circ}$ (Scenario B); without it the residual is essentially the input drift. Right: fraction of seeded perturbations whose final loss reaches the per-spot noise floor (loss $< 1.0$ on the 0.5-px-noise dataset). Naive joint L-BFGS converges on 3/10 Scenario A seeds and 0/10 Scenario B seeds; with the Adam warm-up it converges on all 25 seeds in both scenarios. **(b)** Loss against L-BFGS closure call for the four refinement variants on seed 0. The dashed line marks the per-spot noise floor $\langle (0.5\,\mathrm{px})^2 + (0.5\,\mathrm{px})^2 + (\Delta\omega/\Delta\omega_{\mathrm{step}})^2 \rangle \approx 0.5$; with the Adam warm-up the joint schedule reaches it in both scenarios.

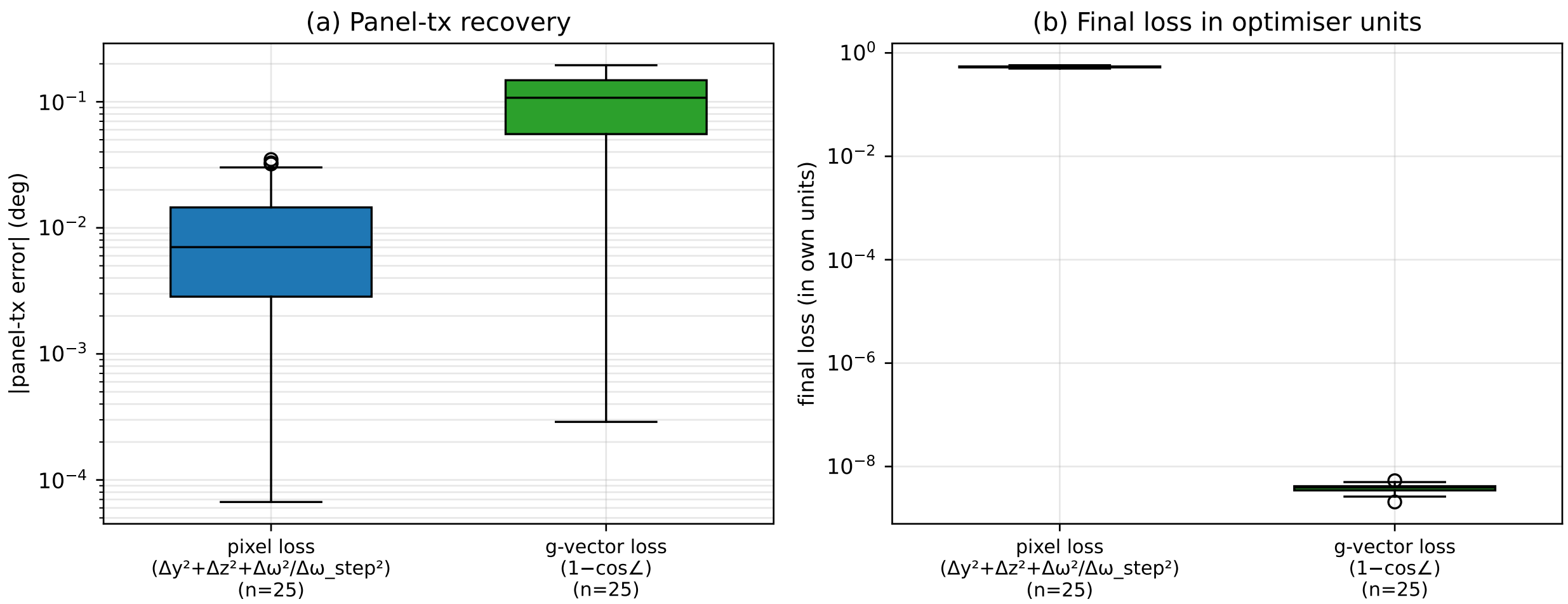

Figure S5: Pixel-loss vs G-vector-loss on Scenario B-joint-fixed-truth (25 seeds each). (a) Panel-$t_x$ residual: pixel loss recovers $t_x$ roughly 10× tighter than G-vector loss. (b) Final loss in the optimiser's own units; the G-vector loss reaches a much smaller numerical value but with no useful signal on $t_x$.

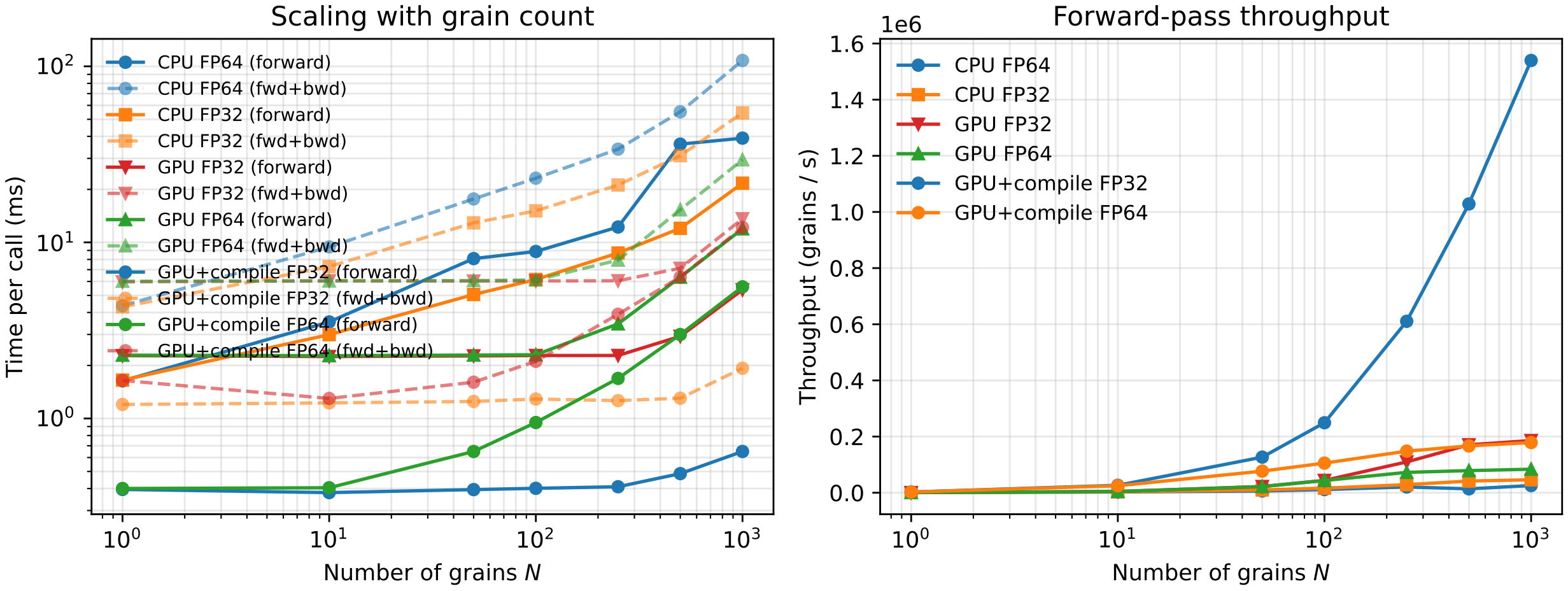

Figure S6: Forward-pass wall-clock vs grain count, FP32 and FP64 on CPU and GPU. The per-grain cost is sub-millisecond on GPU FP32 across the entire range; the FP64 GPU and CPU paths are within 5× of FP32 GPU at typical operating points (a few hundred grains).